\documentclass{aa}
\usepackage{natbib}
\usepackage{amsmath}
\usepackage{graphics}   % standard graphics specifications
\usepackage{graphicx}   % alternative graphics specificati.jpgons
\usepackage{epstopdf}
\usepackage{longtable}   % helps with long table options
\usepackage{lipsum}
\usepackage{booktabs}
\usepackage{array}
\usepackage{hyperref}
\usepackage{afterpage}
\usepackage{supertabular,booktabs}
\usepackage[dvipsnames]{xcolor}
\usepackage{pdflscape}
\def\ga{\,\,\raise0.14em\hbox{$>$}\kern-0.76em\lower0.28em\hbox
{$\sim$}\,\,}
\def\la{\,\,\raise0.14em\hbox{$<$}\kern-0.76em\lower0.28em\hbox
{$\sim$}\,\,}

\begin{document}

\title{The impact of nuclear uncertainties on the p-process nucleosynthesis in Supernovae}
\author{S\'ebastien Martinet\inst{1} \and Stephane Goriely\inst{1} \and Arthur Choplin\inst{1}}

 \institute{Institut d'Astronomie et d'Astrophysique, Universit\'e Libre de Bruxelles (ULB), CP 226, B-1050 Brussels, Belgium\\
 \email: stephane.goriely@ulb.be
 }
 \authorrunning{Martinet et al.}
 \titlerunning{Impact of nuclear uncertainties on the p-process nucleosynthesis}

\date{--}

\abstract
{ The p-process nucleosynthesis is called for to explain the origin of the stable neutron-deficient nuclides heavier than iron that are observed up to now in the solar system exclusively. Our understanding of the p-process nucleosynthesis is still affected by relatively large nuclear uncertainties.  One of the significant uncertainties in determining the ejecta composition stems from the nuclear level densities and photon strength functions entering the calculation of the photodisintegrations of relevance in the high-temperature environment characterizing supernova explosions.
%We present an in-depth study of the nuclear mass uncertainties by varying theoretical nuclear input models that describe the experimentally unknown neutron-rich nuclei. This includes two frameworks for calculating the radiative neutron capture and photoneutron rates and different models for nuclear masses.
}
{We investigate both the model (systematic) and parameter (statistical) uncertainties affecting nuclear level densities and photon strength functions and explore their impact on the p-process nucleosynthesis in type-Ia and type-II supernovae.}
{The impact of correlated model uncertainties affecting nuclear level densities and photon strength functions is estimated by considering different models that are known to provide an accurate description of known related observables. In addition, the uncorrelated uncertainties associated with local variation of model parameters are estimated using a variant of the backward-forward Monte Carlo method to constrain the parameter changes to experimentally known rates before propagating them consistently to the unknown cross sections of neutron-deficient nuclei. 
Both model and parameter uncertainties associated to nuclear level densities and photon strength functions are propagated to the p-process nucleosynthesis taking existing correlations into account. To identify which reactions most strongly control the abundance uncertainties of the p-nuclides, we combine regularized linear-response modeling, stability analysis, and contribution and interaction decompositions.}  
{We find that the uncorrelated parameter uncertainties affecting the photoneutron emission strongly dominate the uncertainty budget. The dominant source of uncertainty arises from the range of local parameter variations still allowed by the present experimental constraints. The main limitation is consequently the lack of sufficiently constraining nuclear data in the relevant neutron-deficient region.  For a large fraction of the p-nuclei, the governing uncertainties are found to be originating from either the photoneutron emission of the p-nucleus itself or the $(\gamma,n)$ reaction on a nearby isotope along the same isotopic chain. 
}  
{Improvements to nuclear models are still crucial in reducing uncertainties in predictions related to the p-process nucleosynthesis. Many of the key reactions identified as dominant drivers correspond to stable or near-stable nuclei, and in several cases to the photodisintegration of the p-nuclei themselves. A substantial fraction of the reactions controlling the p-process abundance uncertainties should, at least in principle, be experimentally accessible through direct measurements.}


\maketitle
\section{Introduction}

Despite major advances over recent decades, the origin of the elements heavier than iron remains debated and is not yet fully resolved \citep[e.g.][]{Arnould20}. The slow (s) and rapid (r) neutron‑capture processes are the two major mechanisms, each contributing roughly half of the origin of trans‑iron nuclides in the Universe.
The s‑process 
\citep[e.g. see the review by][]{Lugaro23}
occurs during the late evolutionary phases of asymptotic giant branch (AGB) stars \citep[main s‑process; e.g.][]{Gallino98,Herwig03,Straniero06,Karakas14} and during core helium burning and shell carbon burning in massive stars \citep[weak s‑process; e.g.][]{Langer89, Prantzos90, The07,Pignatari10,Choplin18, Limongi18}.
In contrast, the r‑process 
\citep[e.g. see the review by][]{Arnould03,Cowan21}
is associated with explosive environments such as neutron‑star mergers \citep[e.g.][]{Goriely11b, Wanajo14, Just15,Just23}, magnetorotational supernovae (SNe) \citep{Winteler12, Nishimura15}, or collapsars \citep{Siegel19b}.
At neutron densities between those characteristic of the s‑ and r‑processes, an intermediate neutron‑capture process (i‑process), first proposed by \citet{Cowan77}, is expected to occur. Its astrophysical site(s) remain a topic of active investigation \citep[see e.g.][ and references therein]{Choplin24,Wiedeking25}.
However, none of these neutron-capture processes can be called for to explain the origin of the 35 neutron-deficient stable isotopes (p-nuclides) of heavy elements beyond iron. These rare species typically occur at levels of about 1\%–0.1\% of their isobars in the Solar System (SoS) abundance distribution 
\citep{Arnould03,Arnould20,Pignatari16}.
To explain the  origin of these p-nuclides, a rather different process need to be invoked, the so-called p-process, thought to result mainly from sequences of photodisintegration reactions ($\gamma$,n), ($\gamma$,p), and ($\gamma$,$\alpha$), acting on pre-existing r- or s-process seed nuclei \citep{Arnould76,Woosley78,Rayet95,Arnould03}. In some cases, p-nuclides may also form via proton captures (p,$\gamma$) or by decay chains from unstable progenitors \citep{Arnould20,Frohlich06}.
Evidence for p-process nucleosynthesis comes not only from abundance patterns in the Sun, but also from extinct radionuclides such as \textsuperscript{92}Nb and \textsuperscript{146}Sm detected in meteoritic materials, and isotopic anomalies in Mo and Xe (notably the Xe-HL component in presolar grains) \citep{Arnould03}. Despite this, the astrophysical origin of p-nuclides remains only partially understood and several key isotopes continue to elude satisfactory reproduction in model predictions.

Among the astrophysical sites considered for the p-process, core-collapse supernovae (CCSNe) are the most extensively studied. The p-process is believed to occur in the O-Ne-rich layers of massive stars during explosive burning \citep{Rayet95,Arnould03,Travaglio18}. While network calculations show that roughly 60\% of p-nuclides are produced in agreement with SoS abundances (within a factor of 3), significant underproduction persists for several isotopes, such as \textsuperscript{92,94}Mo, \textsuperscript{96,98}Ru, \textsuperscript{113}In, \textsuperscript{115}Sn, \textsuperscript{138}La, and \textsuperscript{152}Gd \citep{Rayet95}.
Type Ia supernovae (SNIa) represent another possible site for p-nuclide production, especially under deflagration or delayed detonation  \citep{Travaglio11,Travaglio15}. While these models yield p-process patterns similar to CCSNe, they particularly underproduce \textsuperscript{180m}Ta, as well as \textsuperscript{92,94}Mo, \textsuperscript{96,98}Ru \citep{Arnould03, Travaglio11}. 

It is now well accepted that the p‑process in CCSNe and SNIa offers a potential explanation for the neutron‑deficient nuclides observed in SoS abundances \citep[e.g.][]{Arnould03, Lugaro16}. However, both models systematically underproduce the p‑isotopes $^{92,94}$Mo and $^{96,98}$Ru. This shortfall has motivated the exploration of alternative or additional nucleosynthetic pathways capable of synthesizing these nuclei.
One proposed solution is the so‑called pn‑process—a proton‑poor, neutron‑enhanced variant of the rp‑process—thought to occur during helium detonation \citep{Goriely05}. This scenario involves a sub‑Chandrasekhar‑mass ($M < 1.4,M_\odot$) carbon–oxygen white dwarf that accumulates a helium‑rich surface layer. Another proposed site for producing Mo and Ru p‑nuclides is the proton‑rich, neutrino‑driven wind in CCSNe, where antineutrino absorption in proton‑rich matter creates neutrons that are rapidly captured by neutron‑deficient nuclei \citep[the $\nu p$‑process;][]{Frohlich06}. The proton richness of the neutrino‑driven wind remains, however, highly debated and affected by many modelling uncertainties.  Another possibility is CCSNe from rotating progenitors, that have experienced enhanced s-process nucleosynthesis during their evolution, leading to a modified distribution of p-process seed nuclei at the time of explosion \citep{Choplin22}. Additional astrophysical sites have been proposed, as reviewed in \citet{Arnould03}, including more recently carbon-oxygen shell mergers \citep{Roberti23}.

In addition to astrophysical uncertainties, many key challenges stem from the underlying nuclear physics. Some of the most persistent discrepancies, including the underproduction of Mo and Ru p-isotopes, are believed to originate not only from the astrophysical environment, but also from uncertainties in the reaction rates themselves \citep{Arnould03,Wiedeking24}. For isotopes like \textsuperscript{138}La, the final abundance results from a sensitive balance between production via \textsuperscript{139}La($\gamma$,n) and destruction via \textsuperscript{138}La($\gamma$,n), both of which depend strongly on nuclear structure and input models \citep{Goriely01b,Kheswa15,Kheswa17}. In the case of \textsuperscript{180}Ta, only its long-lived isomer \textsuperscript{180m}Ta is observed in nature. Its survival probability depends on thermal equilibration conditions in stellar environments and is affected by large uncertainties in ($\gamma$,n) rates on both \textsuperscript{180}Ta and \textsuperscript{181}Ta \citep{Goko06,Malatji19}.

Given the extreme scarcity of experimental nuclear data for neutron-deficient, unstable isotopes involved in the p-process, theoretical reaction rates must be computed using the Hauser-Feshbach statistical model \citep{Holmes76,Koning23}. In such models, nuclear-level densities (NLDs) and photon strength functions (PSFs) are among the dominant inputs determining radiative capture cross sections. These inputs, however, are subject to large uncertainties. The situation is particularly intricate for the p-process, where the relevant photodisintegration rates, ($\gamma$,n), ($\gamma$,p), and ($\gamma$,$\alpha$), are not computed directly, but instead derived from the inverse capture rates via detailed balance \citep{Holmes76}. Therefore, the uncertainty in the capture rates directly translates into uncertainty in the photodisintegration channels.

In \citet{Martinet24}, we applied the so-called Backward–Forward Monte Carlo (BFMC) approach to systematically quantify the uncertainties in reaction rates arising from variations in NLD and PSF parameters, and to assess their impact on the resulting i-process abundances in AGB stars. This method filters out non-physical input combinations based on agreement with known experimental Maxwellian-averaged cross sections (MACS) at 30~keV, and propagates the surviving parameter sets to unmeasured reactions of astrophysical relevance. The resulting minimum and maximum rates span a physically meaningful and data-consistent uncertainty band, while preserving correlations between inverse reactions.

In the present work, we extend this framework to the p-process regime. Using the same BFMC methodology reviewed in Sec.~\ref{sec:method}, we compute (n,$\gamma$) rates over a wide range of proton-rich nuclei and derive the corresponding ($\gamma$,n) rates through detailed balance. We additionally compute (p,$\gamma$) and ($\alpha$,$\gamma$) reactions, and obtain ($\gamma$,p) and ($\gamma$,$\alpha$) rates accordingly. To maintain internal consistency, all reaction channels involving the same nucleus are calculated using the same set of NLD and PSF parameters, thereby preserving the expected correlations between them. This allows us to construct correlated Monte Carlo sets of reaction rates, which are then propagated through a full p-process nucleosynthesis network to assess their impact on the final abundances, as detailed in Sec.~\ref{Sect:Impact_p-process}.
This approach enables us to systematically explore in Sec.~\ref{sec:rates} how nuclear input uncertainties propagate into p-nuclide predictions and to identify the most influential reactions, both to prioritize experimental efforts and to refine theoretical models. Final conclusions are drawn in Sec.~\ref{sec:conc}

\section{Method}
\label{sec:method}
Any nuclear input of astrophysical interest is affected by both model and parameter uncertainties. Model uncertainties originate from the intrinsic limitations of the physical assumptions or approximations in a given model. For p-process applications requiring extrapolation away from experimentally known region, it is of prime importance to restrict models to the most accurate ones in reproducing experimental observables, but also to the most fundamental ones based as much as possible on sound and microscopic approaches. In turn, any model rely on adjustable parameters. Parameter uncertainties stem from the incomplete experimental knowledge or theoretical understanding of these values within a given model. Unlike model uncertainties, parameter uncertainties reflect the model defects and the variability in parameter choices that still allow the model to reproduce known experimental data.

Determining model and parameter uncertainties in nuclear astrophysics and propagating them into nucleosynthesis calculations is particularly challenging due to the complex, multidimensional nature of nuclear models and the vast number of adjustable parameters they require, but also the correlation they may embody.
While significant effort has been devoted to characterizing model uncertainties in reaction rate predictions, far less attention has been paid to parameter uncertainties within a given nuclear model. These uncertainties in radiative capture rates predictions arise from local variations of the parameters that govern the NLD and PSF and can have a considerable impact on the predicted rates \citep{Martinet24}. 
When dealing with proton- or $\alpha$-capture reactions, the optical model potential plays a crucial role in determining the reaction rate, especially at energies relevant to the p-process. The associated uncertainty quantification requires different nuclear observables to constrain the optical model parameters than those used for NLDs and PSFs \citep[see e.g.][]{Whitehead21,Pruitt23}.
For this reason, they will not be addressed in the present work but postpone to some future studies.

In this work, we compute radiative neutron, proton and alpha capture and inverse photodisintegration rates using the Hauser-Feshbach code \textsc{TALYS} \citep{Koning23}, adopting a combination of microscopic models for both the NLD and PSF. Specifically, we employ the combinatorial NLDs based on the Hartree-Fock-Bogoliubov (HFB) formalism \citep{Goriely08b}, together with dipole E1 and M1 PSFs obtained from the D1M+QRPA model \citep{Goriely18a}. These choices are motivated by their microscopic consistency and broad applicability across the nuclear chart.

Even within this framework, the predictions are sensitive to a small number of key parameters that capture the leading sources of local uncertainty. For the HFB+comb NLDs, the rate predictions can be systematically modified through two parameters: $\alpha$, which effectively scales the level density, and $\delta$, which introduces an energy shift in the excitation spectrum 
\citep{Koning08}.
Similarly, for the D1M+QRPA PSFs, we account for uncertainties in the centroid energy and width of the giant dipole resonance through two multiplicative parameters, denoted $\delta_E$ and $\delta_\Gamma$, respectively 
\citep[see in particular Fig.~21 of][]{Koning23}. 
These modify the photon strength distributions in a manner consistent with observed variabilities in experimental systematics.

In total, we thus consider four local parameters, two affecting the NLD and two the PSF, which are varied simultaneously to explore the uncertainty space associated with this specific physical model. Importantly, the allowed variation ranges for each parameter are not arbitrarily chosen but constrained by experimental data, as detailed below.

% By propagating this filtered ensemble of NLD and PSF parameter sets through \textsc{TALYS}, we compute correlated $(n,\gamma)$, $(p,\gamma)$, and $(\alpha,\gamma)$ reaction rates for neutron-deficient nuclei involved in the p-process. These rates serve as the basis for evaluating the uncertainty range in the inverse $(\gamma,n)$, $(\gamma,p)$, and $(\gamma,\alpha)$ channels via detailed balance. The resulting uncertainty bands thus reflect the intrinsic variability of the microscopic model under local parameter fluctuations constrained by experimental data.
\subsection{Backward Forward Monte Carlo approach}
To propagate parameter uncertainties in a physically meaningful way, we adopt the BFMC approach developed by \citet{Bauge11}. The method proceeds in two stages. In the first, or “backward” step, combinations of nuclear model parameters, here those controlling the NLD and PSF, are sampled and filtered by comparing the resulting reaction rates against available experimental data, such as MACS measurements. Only parameter sets reproducing the data within uncertainties are retained, thereby constraining the variations to empirically supported regions of the parameter space.

In the second, or "forward" step, these filtered sets are used to compute reaction rates for nuclei where no experimental information is available. By systematically applying these constrained parameter combinations to the calculation of $(n,\gamma)$, $(p,\gamma)$, and $(\alpha,\gamma)$ rates, and deriving the corresponding photodisintegration rates via detailed balance, we obtain physically consistent uncertainty bands that account for model parameter variability while preserving correlations across channels.

In the present work, we adopt exactly the same BFMC method as detailed in our previous \textit{i}-process study of \cite{Martinet24}. In this work, 
we implemented a BFMC approach in which thousands of parameter combinations were sampled, and only those reproducing known MACS within experimental uncertainties were retained. In the present study, we re-use the subset of BFMC-filtered parameter sets obtained in that work, ensuring that all variations are physically consistent and empirically grounded. 

\subsection{Nuclear physics input and parameter uncertainties}
\label{sec:nuclear_uncertainties}

% As in our previous work on the \textit{i}-process \citep{martinet2024}, we rely on a physically motivated approach to quantify the uncertainties in theoretical reaction rates, grounded in the systematic propagation of nuclear model parameters constrained against available experimental data. In the context of the p-process, where photodisintegration reactions dominate the nucleosynthetic flows, a key source of uncertainty stems from the theoretical evaluation of $(\gamma,\textit{x})$ reaction rates—namely $(\gamma,n)$, $(\gamma,p)$, and $(\gamma,\alpha)$—which are computed through detailed balance from their respective capture counterparts.

In this study, we consider a reaction network comprising radiative neutron, proton and alpha captures as well as their reverse photodisintegration reactions on about 2000 neutron-deficient isotopes up to Po. Almost all of the corresponding rates are experimentally unconstrained. These channels are essential because the nucleosynthesis path in the p-process depends sensitively on the competition between $(\gamma,n)$, $(\gamma,p)$, and $(\gamma,\alpha)$ photodisintegrations, which are all computed from their inverse reactions using detailed balance. Thus, for each of the filtered BFMC parameter sets, we consistently compute correlated $(n,\gamma)$, $(p,\gamma)$, and $(\alpha,\gamma)$ rates across the full network of experimentally unconstrained reactions. This allows us to preserve the physical correlations between rate predictions originating from a common nuclear input configuration, i.e. a coherent NLD or PSF for a given nucleus.

To quantify the range of uncertainties, we evaluate the maximum and minimum values of each $(n,\gamma)$ rate across all parameter sets obtained from the BFMC method for each target nucleus. We then identify, for each nucleus, the parameter sets that produce these extrema. Importantly, we associate to each extreme $(n,\gamma)$ rate the corresponding $(p,\gamma)$ and $(\alpha,\gamma)$ rates computed with the same nuclear input set, thereby ensuring that all rates remain physically correlated across the three channels. This procedure results in a set of correlated, extreme-case scenarios that reflect the maximal spread in nuclear model predictions while respecting internal consistency.

\begin{figure*}
    \centering
    \includegraphics[width=0.9\textwidth]{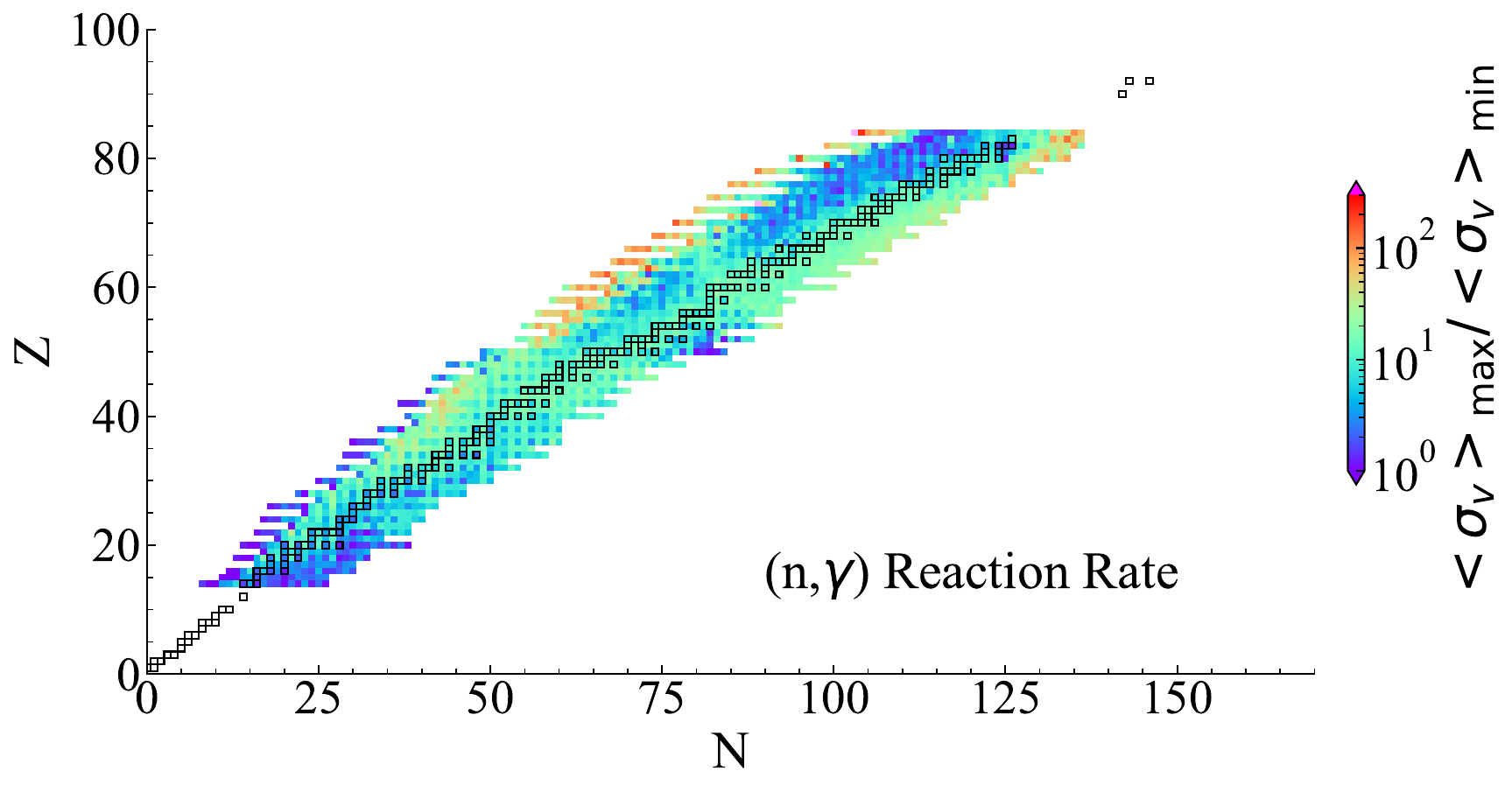}
\caption{Representation in the ($N$,$Z$) plane of the nuclear parameter uncertainties on the $(n,\gamma)$ reaction rates for the p-process. The color-code depicts the ratio between the maximum and minimum rate obtained for each nuclei with the BFMC method} 
\label{fig:NZ_max_min_rate_p_pro_n}
\end{figure*}

Figure \ref{fig:NZ_max_min_rate_p_pro_n} shows the range of nuclear parameter uncertainties of $(n,\gamma)$ rate for our 2000 nuclei in the p-process network. While straying further away from the stability zone tends to increase the uncertainty range, we can see that the impact of the NLD and PSF uncertainties on the proton-rich side above $Z=60$ is actually smaller than on the neutron-rich side. 
Indeed, for neutron-deficient nuclei, the radiative neutron capture rate becomes rapidly smaller than the competing  $(n,p)$ reaction rate. Additionally, with an increasing neutron separation energy in the neutron-deficient side, the $(n,\gamma)$ rate becomes sensitive to the giant dipole resonance region rather than to its uncertain tail. 

Figure \ref{fig:NZ_max_min_rate_p_pro_a_sub} and \ref{fig:NZ_max_min_rate_p_pro_p_sub} show the corresponding $(p,\gamma)$ and $(\alpha,\gamma)$ rates associated to each extreme $(n,\gamma)$ rates, meaning computed with the same nuclear input set, ensuring correlation between the three channels. We can see that the uncertainty in the proton-rich side is very limited due to the fact that the NLD and PSF have a limited impact on the $(p,\gamma)$ and $(\alpha,\gamma)$ rates. 
The $(\gamma,p)$ and $(\gamma,\alpha)$ channels play, however, an important role in shaping the the p-process nuclear flow by photodisintegrating material to lighter $Z$-elements \citep{Rayet95,Arnould03}.
This suggests the need to explore in future works the impact of the parameter uncertainties associated with the proton- and $\alpha$ optical potentials on the rates. 

In addition to this global analysis of joint variation across channels, we also examine the effect of varying each reaction type independently. That is, for each nucleus, we separately determine the maximum and minimum of the $(n,\gamma)$, $(p,\gamma)$, and $(\alpha,\gamma)$ rates in isolation, while keeping the other channels fixed at their nominal values. This allows us to assess the specific role each channel plays in shaping the p-process abundances and to isolate the individual contributions to the total (NLD- and PSF-related) uncertainty budget.

This multi-channel extension of the BFMC framework thus enables a comprehensive and physically consistent evaluation of the nuclear uncertainties affecting p-process nucleosynthesis, rooted in the same methodology that was successfully applied to the \textit{i}-process regime. The resulting set of correlated rates serves as the input for our sensitivity and propagation studies described in the following section.

\section{Impact on the p-process nucleosynthesis}
\label{Sect:Impact_p-process}

\subsection{Supernovae simulations}
Various scenarios have been proposed to account for the bulk p-nuclide content of the SoS. 
The p-nuclides are primarily produced during the final explosion of a massive star ($M \ga 10~M_\odot$) as a CCSN or during the pre-explosive oxygen burning episode \citep{Arnould03}. The p-process can occur in the O-Ne layers of massive stars explosively heated to peak temperatures ranging between 1.7 and $3.3 \times 10^9$~K. The seeds for the p-process consist of heavy nuclei either inherited at stellar birth or synthesized by the s-process during earlier hydrostatic burning phases. SNIa have also been proposed as potential sites for the p-process. P-process nucleosynthesis, possibly accompanying the deflagration or delayed detonation regimes, has been predominantly studied in 1D simulations, revealing overabundances similar to CCSN models \citep{Arnould03,Travaglio15}. However, predicted p-nuclide yields from SNIa suffer from considerable uncertainties affecting the adopted explosion models and the s-seed distributions, with detailed information on the composition of the material transferred to the white dwarf before the explosion being unavailable.

\begin{figure*}
    \centering
    \includegraphics[width=0.9\textwidth]{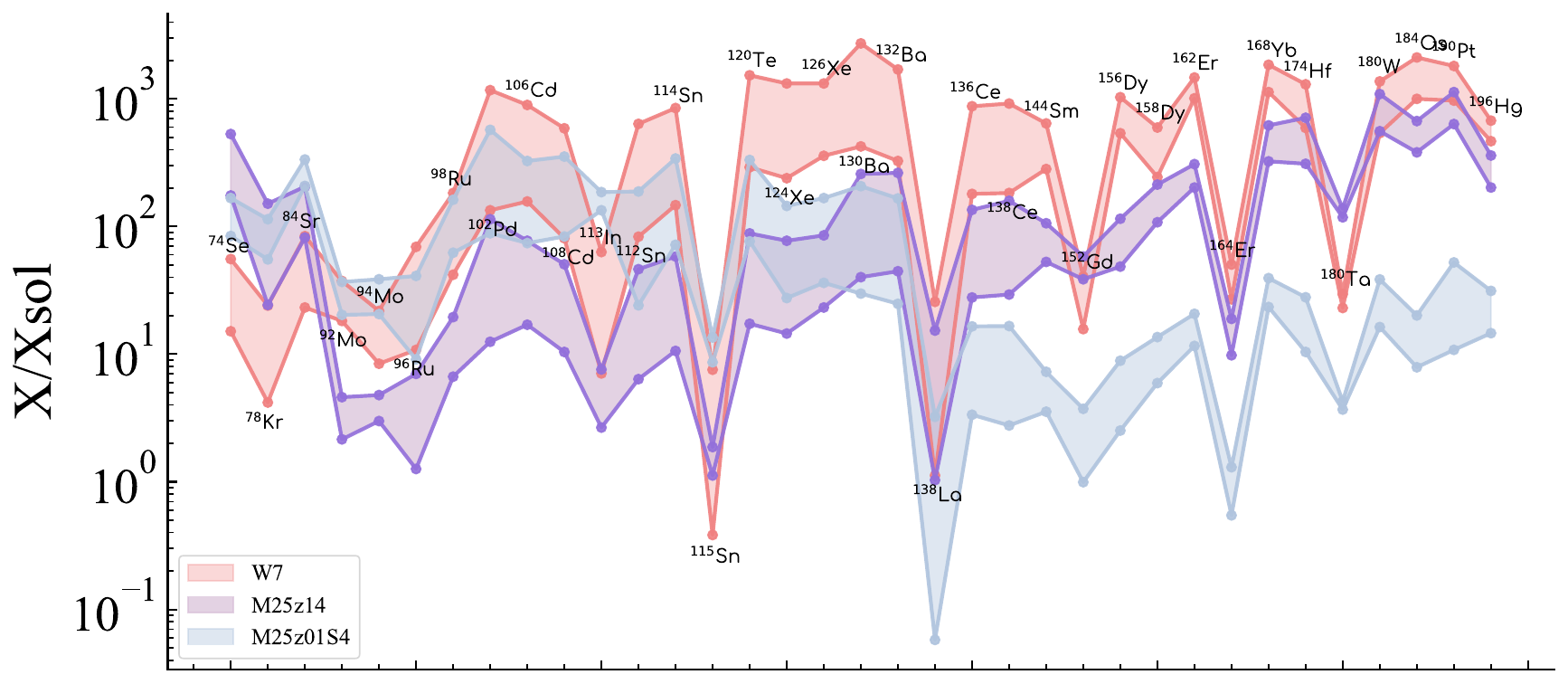}
\caption{Final p-nuclide abundances obtained for the three astrophysical sites considered in this work: the W7 Type Ia supernova model, a $25\,M_\odot$ massive star at solar metallicity, and a rotating $25\,M_\odot$ model at low metallicity. The shaded regions represent the abundance uncertainty ranges resulting from the propagation of nuclear parameter uncertainties.}
\label{fig:abund_triple_site}
\end{figure*}

To study the impact of nuclear uncertainties on p-process nucleosynthesis, we consider three different SN models, namely
\begin{itemize}
    \item an exploding non-rotating massive star of $25~M_\odot$ and solar metallicity (hereafter M25z14) 
    \citep{Rayet95}.
    \item an exploding rotating massive star with $25~M_\odot$ and a metallicity of $Z = 10^{-3}$ (hereafter M25z01S4) that have undergone an enhanced s-process nucleosynthesis during their life through rotational mixing \citep{Choplin22}. We adopt an initial rotation velocity of $v_{\rm ini}/v_{\rm crit} = 0.4$, where $v_{\rm crit}$ is the critical velocity at which the gravitational acceleration is compensated by the centrifugal force.
    \item a detonating SNIa, described by the W7 model of \citep{Nomoto84}. 
\end{itemize}
The resulting overproduction factors (with respect to the solar composition) are shown in Fig.~\ref{fig:abund_triple_site} for the 35 p-nuclides. The shaded areas represent the abundance uncertainties resulting from the propagation of the nuclear parameter uncertainties, as discussed below.

%The NLDs and PSFs are seen to have a relatively small impact on the overall p-distribution. Indeed, the photodisintegration rates are estimated from the reverse radiative capture rates on the basis of the detailed balance principle and, as seen in Fig.~\ref{fig_signpa}, neutron-deficient nuclei have  p- and $\alpha$-captures rates (hence photo-rates) rather insensitive to NLDs and PSFs. Deviations in the predicted yields essentially come from the more sensitive photo-neutron rates but their impact on the p-process yields remains marginal.

\subsection{Propagating parameter uncertainties}

To propagate the parameter uncertainties to the p-process abundance predictions, we adopt the same strategy as in \citet{Martinet24}. In that work, a large number of nuclear sets was constructed by randomly combining, for each reaction, the minimum and maximum rates obtained from the BFMC analysis, and these sets were then propagated through full nucleosynthesis calculations. The resulting abundance distributions were used to estimate uncertainty intervals in a statistically robust way. A dedicated convergence study showed that $N=30$ random realizations were sufficient to capture the dominant part of the propagated abundance spread.

The same procedure as in \citet{Martinet24} is applied here, but extended to the p-process network. For each Monte Carlo realization, we generate a full set of reaction rates by assigning to each nucleus one of the extreme BFMC rate combinations retained in Sec.~\ref{sec:nuclear_uncertainties}. In contrast to the smaller network considered in \citet{Martinet24}, the present application involves a substantially larger set of reactions relevant to the p-process, including not only $(n,\gamma)$ rates and their inverse $(\gamma,n)$ channels, but also $(p,\gamma)$ and $(\alpha,\gamma)$ reactions together with their corresponding photodisintegration rates. For a given nucleus, these different channels are kept correlated, since they are all derived from the same retained BFMC parameter set, whereas the choices made for different nuclei are treated as independent. In this way, the propagation preserves the physically meaningful correlations between reaction channels attached to the same nucleus while still exploring the broader uncertainty space of the full network.

% Each of the $N=30$ generated nuclear sets is then used as input to a full p-process nucleosynthesis calculation. The ensemble of resulting abundance patterns provides, for each p-nuclide, a distribution of final abundances from which uncertainty intervals can be extracted. As in \citet{Martinet2024}, we characterize these distributions using percentile-based ranges rather than absolute extrema in order to reduce the influence of rare outliers or isolated numerical artifacts. The propagated abundance intervals presented below therefore reflect the statistical spread induced by local nuclear-parameter variations within the BFMC framework, now applied to the more extended and multi-channel p-process network.
% \subsection{Isolated impact of p, n and $\alpha$ channels}
% Figure \ref{fig:W7_abund_p_n_a_isolated} shows the impact of each channel uncertainty isolated against the final abundances taking into account the correlations in-between the three channels.
\subsection{Isolated impact of $p$, $n$ and $\alpha$ channels}
\label{sec:isolimp}

\begin{figure*}
    \centering
    \includegraphics[width=0.9\textwidth]{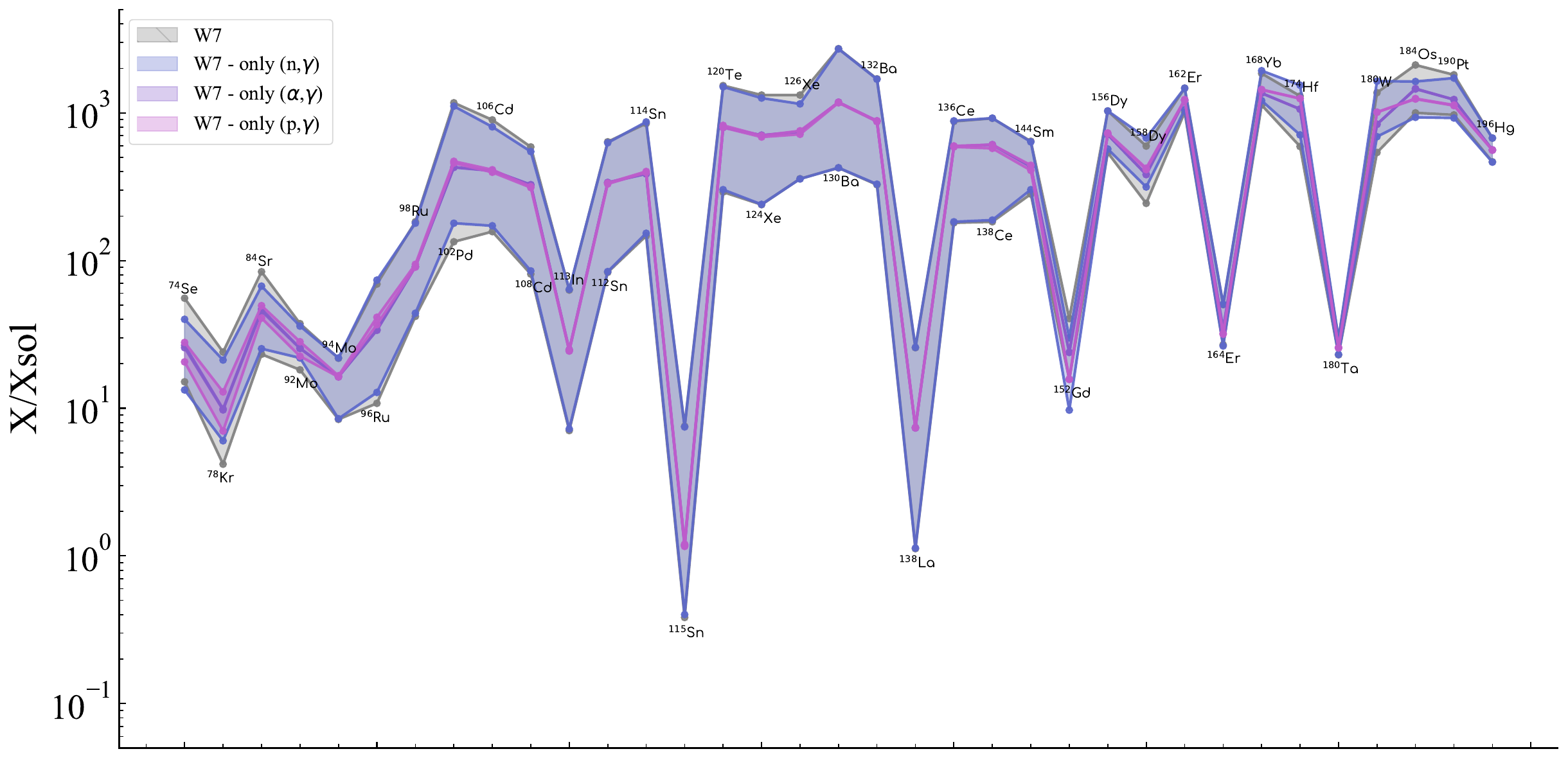}
\caption{Final abundance uncertainty in the W7 model with all three channels correlated and with each isolated channels} 
\label{fig:W7_abund_p_n_a_isolated}
\end{figure*}

Figure~\ref{fig:W7_abund_p_n_a_isolated} illustrates the impact of the uncertainties associated with each reaction channel when varied independently, while preserving the correlations between the $(n,\gamma)$, $(p,\gamma)$ and $(\alpha,\gamma)$ rates attached to a given nucleus. This decomposition allows us to isolate the relative contribution of each channel to the total abundance uncertainty.

The results clearly show that the dominant contribution arises from the $(n,\gamma)$ reactions. When only the neutron-capture channel is allowed to vary within the BFMC uncertainty limits, the resulting spread in the final p-nuclide abundances already accounts for most of the total uncertainty observed when all channels are varied simultaneously. In contrast, when only the $(p,\gamma)$ or $(\alpha,\gamma)$ rates are perturbed, the resulting abundance variations remain very small across the entire network.

This behaviour reflects the different sensitivities of the reaction rates to the nuclear inputs explored in the BFMC framework. As discussed in Sec.~\ref{sec:nuclear_uncertainties}, the uncertainties associated with the NLDs and PSFs have a significantly stronger impact on neutron-capture rates than on proton- or $\alpha$-capture reactions in the proton-rich region relevant to the p-process. As a consequence, the uncertainty budget of the network is largely controlled by the $(n,\gamma)$ channel.

This result also explains why the uncertainty band obtained when all three channels are varied simultaneously is very similar to that obtained when only the $(n,\gamma)$ rates are perturbed. In practice, the additional variations introduced by the $(p,\gamma)$ and $(\alpha,\gamma)$ channels have only a marginal effect on the final abundance spread.

Based on this observation, we adopt in the following sections a simplified strategy to explore the extreme nuclear scenarios. For each nucleus, we identify the BFMC parameter sets that produce the minimum and maximum $(n,\gamma)$ rates and we associate to these extrema the corresponding $(p,\gamma)$ and $(\alpha,\gamma)$ rates obtained with the same nuclear input configuration. This approach allows us to probe the full abundance uncertainty interval while preserving the physically consistent correlations between the different reaction channels.

\subsection{Impact on different p-process sites}

Figure~\ref{fig:abund_triple_site} compares the propagated abundance uncertainties obtained for the three astrophysical scenarios considered in this work: the W7 SNIa model, a $25\,M_\odot$ massive star at solar metallicity, and a rotating $25\,M_\odot$ massive star at low metallicity. Although the overall nucleosynthetic patterns differ between these environments, the magnitude of the nuclear uncertainty remains remarkably similar across all three cases.

In all three models, the propagated nuclear uncertainties lead to abundance variations of comparable magnitude. When averaged over the p-nuclides, the typical uncertainty interval is about $0.7$ dex between the maximum and minimum predicted abundances, corresponding to roughly a factor of five in linear scale. This similar uncertainty level across the three environments indicates that the nuclear parameter variations explored in this work affect the p-process yields in a broadly comparable manner, independently of the specific astrophysical site.

While the layers in all three models are explosively heated to peak temperatures ranging between 1.7 and 3.3~GK, 
the main differences between the models arise from the underlying astrophysical conditions, which determine the seed distributions available for photodisintegration. The W7 model and the solar-metallicity massive star exhibit a relatively efficient production of the heaviest p-nuclides, reflecting the larger availability of heavy seed nuclei inherited from prior stellar nucleosynthesis. In contrast, the low-metallicity rotating model favors the synthesis of light p-nuclides, reflecting the efficient production of light s-process elements during hydrostatic evolution. As a result, the abundance distribution in this case is shifted towards the lower-mass region of the p-process, with significant production mainly up to nuclei around the Ba region.

\subsection{Systematic vs statistical uncertainties}

\begin{figure*}
    \centering
    \includegraphics[width=0.9\textwidth]{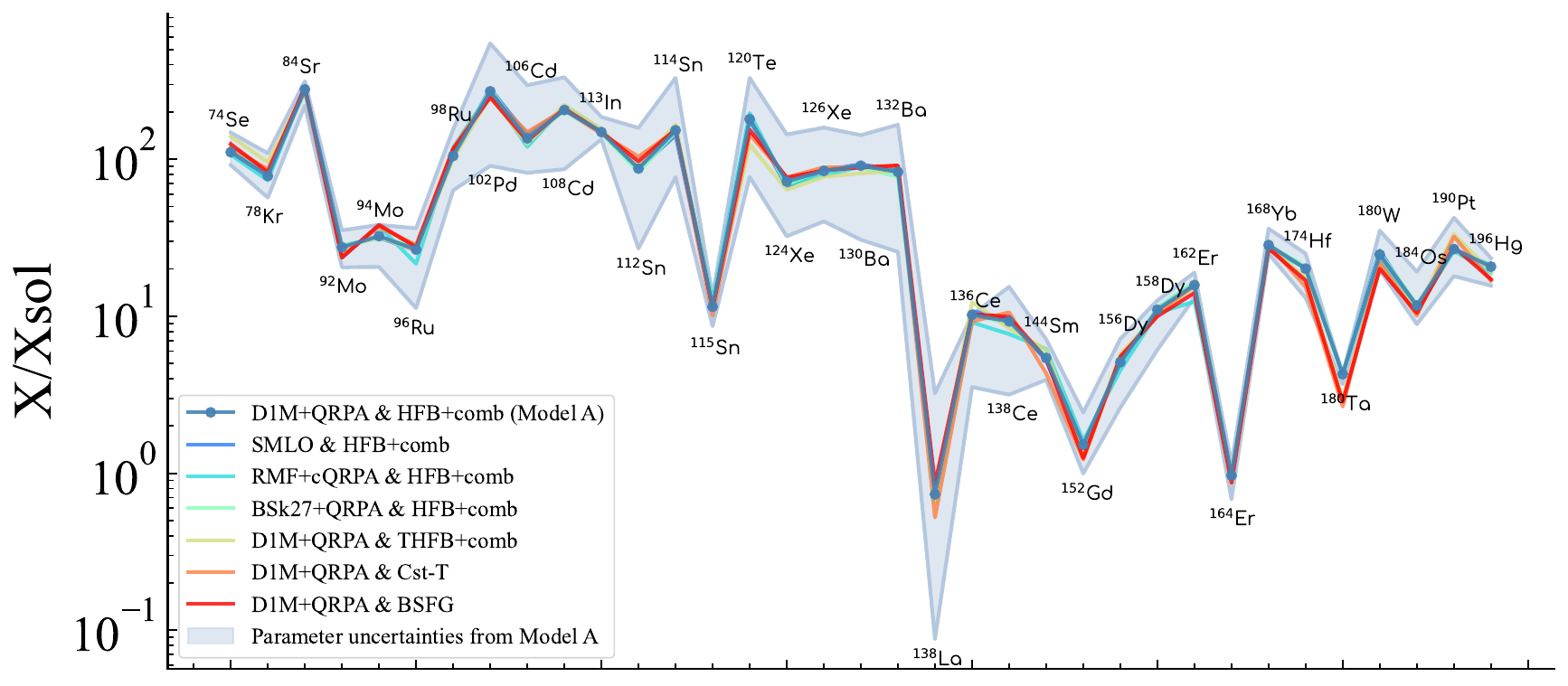}
\caption{Comparison between statistical (parameter) and systematic (model) uncertainties for $(n,\gamma)$ reaction rates relevant to the p-process in the M25z01S4 model. The shaded region corresponds to the parameter uncertainty band obtained with the BFMC approach for the reference model A (D1M+QRPA \& HFB+comb). The colored lines represent the rates predicted by different combinations of PSF and NLD models.}
\label{fig:model_vs_parameters}
\end{figure*}

To assess the relative importance of systematic model uncertainties and statistical parameter uncertainties, we compare the uncertainty band obtained with the BFMC approach for our reference nuclear model (Model A which adopts the D1M+QRPA PSF \& HFB+comb NLD) with the reaction rates predicted by several alternative combinations of PSF and NLD models using the default recommended parameters. More specifically, in addition to the the D1M+QRPA model, we consider the following PSF models: the Simple Modified Lorentzian (SMLO) \citep{Plujko18}, the relativistic mean-field plus continuum QRPA (RMF+cQRPA) of \citet{Daoutidis12a} and the Skyrme plus QRPA (BSk27+QRPA) of \cite{Xu21}. Concerning the NLD models, as an alternative to the HFB+comb model, we also consider the constant-temperature (Cst-T) and Back-Shifted Fermi Gas (BSFG) models of \citet{Koning08}. The combination of PSF and NLD models applied to the p-process simulations keeping the model correlations are listed in the legend of Fig.~\ref{fig:model_vs_parameters}.

The shaded region in Fig.~\ref{fig:model_vs_parameters} represents the range of p-nuclei overabundances resulting from the parameter uncertainty obtained by varying the NLD and PSF parameters within the BFMC framework, while the curves for a given PSF and NLD combination show the rates predicted by the different nuclear models without parameter variation. The comparison reveals that the different models produce remarkably similar rate predictions. In practice, the variations between the different model combinations remain very small compared to the BFMC parameter uncertainty band. When averaged over the network, the spread between models corresponds to only a few percent, typically less than $\sim10\%$ for individual nuclei. 

In contrast, the parameter uncertainties obtained from the BFMC approach are much larger, leading to typical variations of about a factor of five between the maximum and minimum rates. The difference is particularly visible for nuclei in the region extending roughly from Rb to La, where the parameter uncertainty band clearly dominates over the systematic model variations.

This comparison indicates that, within the class of modern microscopic NLD and PSF models considered here, the systematic uncertainties associated with the choice of the nuclear model are relatively small. Although the different theoretical models tend to predict relatively different rates, the correlation imposed by the model when propagating the uncertainties into the p-process simulations strongly reduces the possible spread in the abundance calculations. Instead, the dominant source of uncertainty arises from the local variations of model parameters allowed by the current experimental constraints which are propagated into the p-process simulation in an uncorrelated way. 

This result highlights the importance of improving experimental constraints on the nuclear properties that govern radiative capture reactions in the proton-rich region relevant to the p-process. Additional measurements would help reduce the parameter space explored by the BFMC method and therefore narrow the resulting uncertainty bands on the reaction rates.

In the following section, we identify key rates that drive the uncertainty of the p-process nucleosynthesis for each p-nuclei.

{
\section{Determining important rates}
\label{sec:rates}

\subsection{Statistical identification of key reactions}

The abundance variations obtained from the Monte Carlo rate sets described in Sec.~\ref{sec:method} were used to identify the reaction rates that control the final uncertainties of each p-nucleus. For each target p-nucleus, the input variables are the adopted minimum or maximum values of the reaction rates in a given simulation, while the response variable is the corresponding final abundance of the target nucleus.

For each target p-nucleus, the analysis was performed within local windows in the $(N,Z)$ plane centred on the target nucleus. These windows define the subset of neighbouring reactions included as candidate predictors of the abundance variation. Seven window sizes were considered, starting from a compact region around the target nucleus and progressively expanding to a nearly global network selection. The final driver classification was then based on reactions recovered consistently across the different window sizes, rather than on a single arbitrary window definition. Technical details are given in Appendix~\ref{app:ml_details}.

The statistical workflow consists of three main steps. First, a conservative ElasticNet screening is used only to remove reaction rates with negligible influence on the abundance variations. The retained candidates are then analysed with a Random Forest regressor, and their contributions are decomposed into direct and coupled effects using SHAP values. Finally, reactions are classified in the two-dimensional space defined by their normalized main effect, which measures the direct contribution to the abundance variation, and their normalized interaction strength, which measures their contribution through coupled effects with other reactions.

Reaction rates that exhibit simultaneously large direct and interaction contributions are identified as \textit{core drivers}. This definition selects reactions that not only affect the abundance of a given p-nucleus individually, but also participate significantly in the coupled response of the reaction network. The resulting classification is therefore not based on a single importance metric, but on the simultaneous contribution of direct and interaction effects.

\begin{figure*}
\centering
\includegraphics[width=0.7\textwidth]{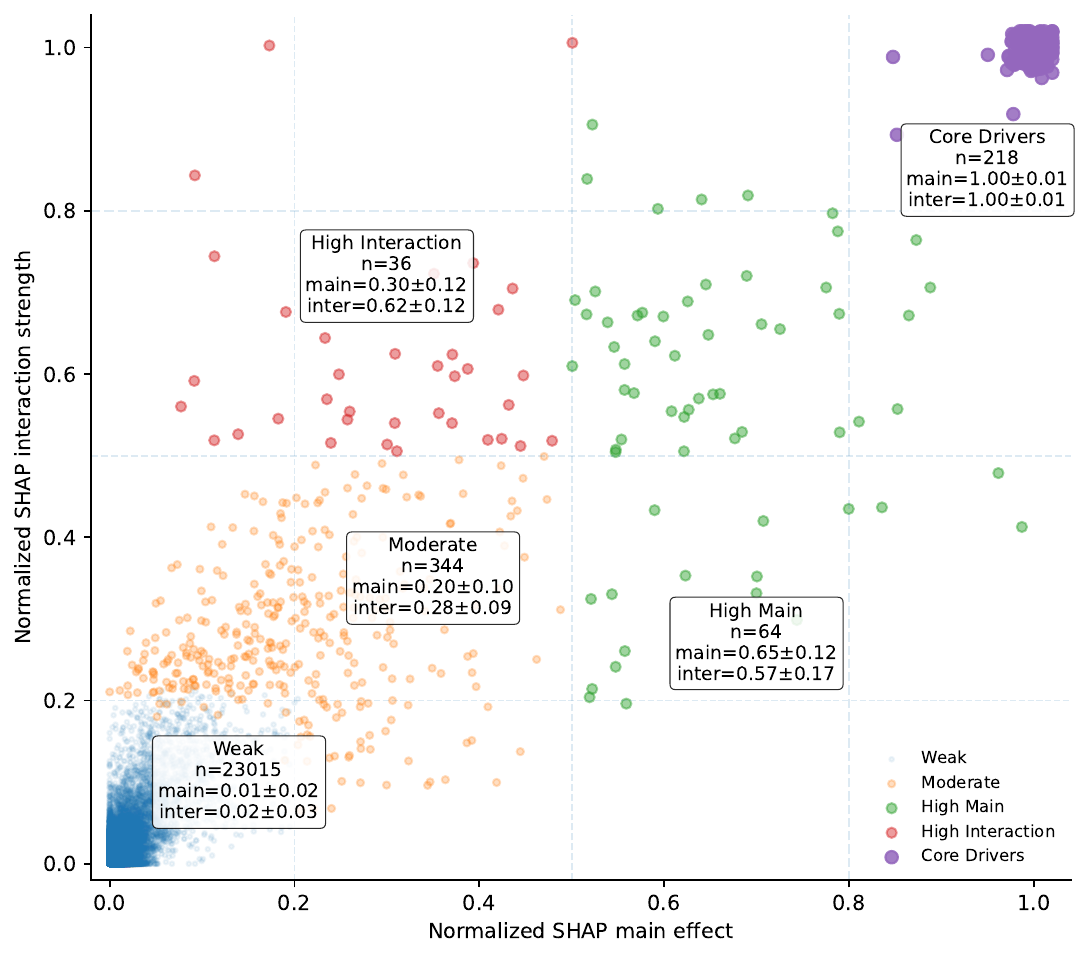}
\caption{
Distribution of all reaction rates in the normalized SHAP main-effect ($M_i$) versus interaction-strength ($I_i$) plane. The main effect quantifies the average direct contribution of a reaction rate to the abundance variation, while the interaction strength measures its contribution through coupled effects with other reaction rates. The thresholds used to define the Core Driver, High Main, High Interaction, and Weak categories are indicated by the dashed lines. Core Drivers occupy the upper-right quadrant and correspond to reactions exhibiting both strong direct and interaction contributions. The number $n$ of reaction rates and the mean $\pm$ standard deviation of the SHAP metrics within each category are reported in the corresponding quadrant.
}
\label{fig:shap_quadrant_category_stats}
\end{figure*}

Figure~\ref{fig:shap_quadrant_category_stats} validates this classification by showing the global distribution of all reaction rates, target nuclei, and analysis windows in the direct-effect versus interaction-effect plane. The core-driver population occupies a statistically distinct region, well separated from the large population of weak contributors. This separation indicates that the selected reactions are not marginal members of a continuous distribution, but correspond to a well-defined subset of reactions that dominate the propagation of abundance uncertainties.

\subsection{Visualization of the core drivers in the $(N,Z)$ plane}
\label{sec:vis}

The core drivers identified by the statistical procedure are represented in the $(N,Z)$ plane in Figs.~\ref{fig:NZ_focus} and \ref{fig:NZ_map_core_drivers_average_vs_top}. Figure~\ref{fig:NZ_focus} first shows a zoomed region around three light p-nuclides, $^{74}$Se, $^{78}$Kr, and $^{84}$Sr, in order to illustrate how the graphical representation should be read. The corresponding full map for the W7 model is then shown in Fig.~\ref{fig:NZ_map_core_drivers_average_vs_top}.

In this representation, each p-nuclide is shown as a colored square, with a distinct color assigned to each isotope. The photodisintegration rates identified as core drivers are represented by circles located at the position of the nucleus involved in the reaction. Each circle has the same color as the p-nucleus whose uncertainty it controls, and a line connects the driver reaction to the affected p-nucleus. The sign placed along the connecting line indicates whether the reaction rate is correlated or anti-correlated with the final abundance. A positive sign denotes that increasing the rate increases the final abundance, while a negative sign indicates that increasing the rate decreases it.

The examples shown in Fig.~\ref{fig:NZ_focus} illustrate the local nature of many of the identified sensitivities. For $^{74}$Se, the abundance uncertainty is mainly linked to the neighbouring reaction $^{76}$Se$(\gamma,n)$, which feeds the p-nucleus through the photoneutron sequence. Similarly, $^{78}$Kr is affected by nearby photoneutron reactions along the Kr isotopic chain, including the feeding reaction from $^{80}$Kr and the direct destruction channel $^{78}$Kr$(\gamma,n)$. The case of $^{84}$Sr provides another example of a driver located close to the p-nucleus itself, showing that the dominant uncertainty propagation often remains confined to the immediate isotopic neighbourhood of the target nucleus.

The outer ring surrounding each driver circle indicates whether the reaction is identified in the ejecta-averaged analysis, in the peak-production-layer analysis, or in both. Drivers recovered by both approaches correspond to particularly robust sensitivities, whereas drivers found only in one of the two analyses reflect a stronger dependence on the adopted layer selection. The grey symbols in the background mark the independently determined photodisintegration branching regions, where $(\gamma,n)$ reactions compete with $(\gamma,p)$ or $(\gamma,\alpha)$ channels over the relevant p-process temperature range. The proximity between the identified drivers and these branching regions provides a useful physical check on the statistical classification.

\begin{figure*}
\centering
\includegraphics[width=0.7\textwidth]{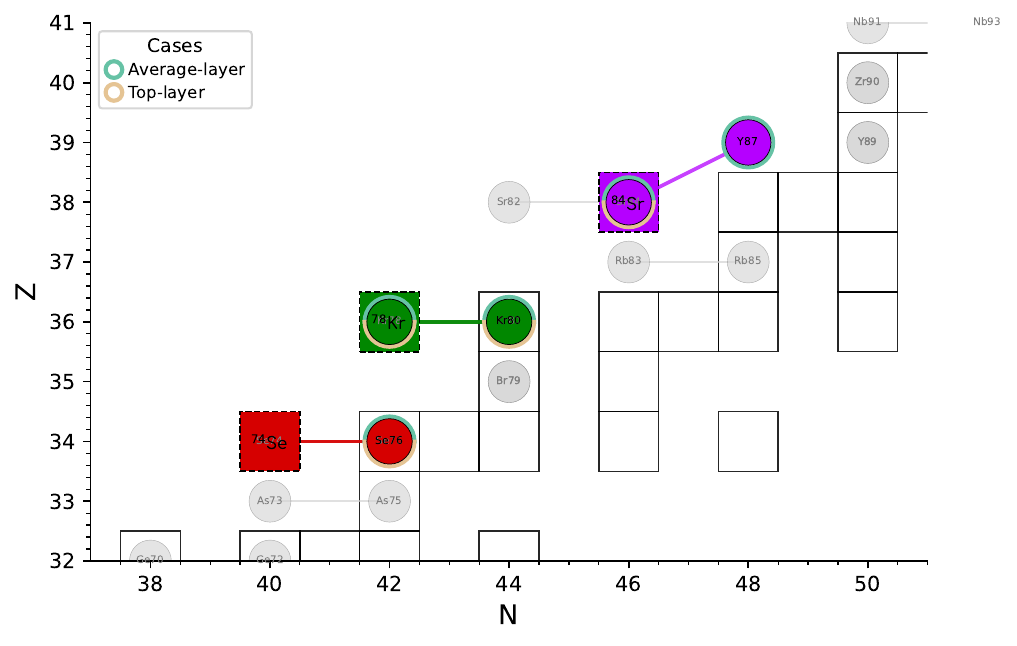}
\caption{
Zoomed view of the core-driver representation in the $(N,Z)$ plane for three light p-nuclides in the W7 model: $^{74}$Se, $^{78}$Kr, and $^{84}$Sr. Colored squares mark the p-nuclides, while circles indicate the photodisintegration reactions identified as core drivers of their abundance uncertainties. Each driver is connected to the p-nucleus whose abundance it affects, and the sign along the connection indicates whether the rate is correlated or anti-correlated with the final abundance. The outer ring indicates whether the driver is recovered in the ejecta-averaged analysis, in the peak-production-layer analysis, or in both. Grey symbols show the independently determined photodisintegration branching regions for temperatures between $T=1.5$ and $3.4\times 10^9$~K. This zoomed panel illustrates how the global maps in Figs.~\ref{fig:NZ_map_core_drivers_average_vs_top} and \ref{fig:NZ_triple_site_reactions} should be read. An interactive version of the nuclear chart is available online at
\href{https://sebastienmartinet.github.io/NZ_driver_map}{https://sebastienmartinet.github.io/NZ\_driver\_map}.
}
\label{fig:NZ_focus}
\end{figure*}

The full W7 map is shown in Fig.~\ref{fig:NZ_map_core_drivers_average_vs_top}. It extends the same representation to all p-nuclides considered in the analysis and provides a global view of the reactions controlling the abundance uncertainties in the W7 model.

\begin{figure*}
\centering
\includegraphics[width=\linewidth, height=0.6\linewidth]{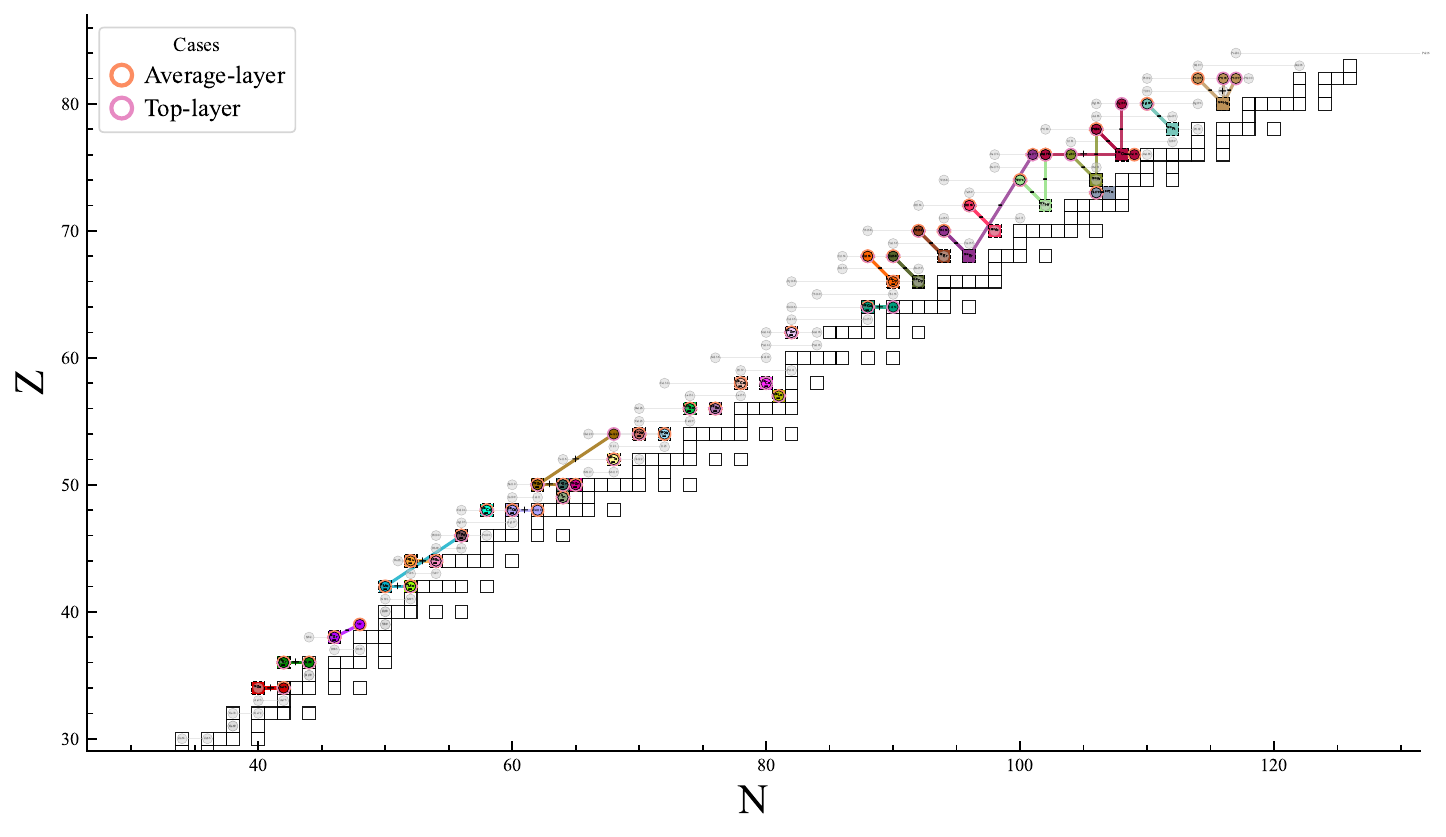}
\caption{
Same graphical representation as in Fig.~\ref{fig:NZ_focus}, but extended to all p-nuclides in the W7 model. Colored squares denote p-nuclides and circles indicate the reaction rates identified as core drivers of their abundance uncertainties. The sign along each connection indicates whether the rate is correlated or anti-correlated with the final abundance. The outer ring distinguishes drivers recovered in the ejecta-averaged analysis, in the peak-production-layer analysis, or in both. An interactive version of the nuclear chart is available online at
\href{https://sebastienmartinet.github.io/NZ_driver_map}{https://sebastienmartinet.github.io/NZ\_driver\_map}.
}

\label{fig:NZ_map_core_drivers_average_vs_top}
\end{figure*}

Two complementary abundance datasets were used to construct these maps. In the first, the abundances were averaged over all mass zones contributing to the p-process production. This `averaged-layer'' analysis provides a global view of the integrated ejecta composition. In the second, the analysis was restricted to the layer where the production of the considered p-nucleus is maximal. This `top-layer'' analysis isolates the thermodynamic conditions most favourable to the production of that isotope. Comparing the two approaches allows us to distinguish reactions that control the global ejecta-integrated abundance from those that are important mainly under the most favourable local production conditions.

For reference, Figs.~\ref{fig:NZ_focus} and \ref{fig:NZ_map_core_drivers_average_vs_top} also show the photodisintegration branching regions obtained independently from the rates uncertainties. These regions correspond to nuclei for which the $(\gamma,n)$ rate becomes comparable to competing $(\gamma,p)$ or $(\gamma,\alpha)$ channels over the relevant p-process temperature range, namely $T=1.5$--$3.4 \times 10^9$~K. The fact that most statistically identified core drivers lie close to these branching regions provides a physical validation of the method: the reactions selected by the statistical analysis are located where variations in the rates are expected to redirect the reaction flow most efficiently.

\subsection{Physical interpretation of the driver distribution}

Several systematic trends emerge from the distribution of core drivers in the $(N,Z)$ plane. For a large fraction of the p-nuclides, especially from $^{74}$Se to approximately $^{152}$Gd, the dominant driver corresponds either to the photoneutron emission of the p-nucleus itself or to the $(\gamma,n)$ reaction on a neighbouring isotope along the same isotopic chain. In practice, the abundance uncertainty of a given p-nucleus is often controlled by the balance between its direct destruction and its feeding by sequential photoneutron emission from heavier isotopes.

This behaviour reflects the dominant role of $(\gamma,n)$ sequences in the classical p-process. During the high-temperature phase, heavy seed nuclei are driven towards the neutron-deficient side by successive photoneutron emissions. The final abundance of a p-nucleus is therefore largely determined by the competition between the reactions that feed it from heavier isotopes and the reactions that destroy it through further photodisintegration.

For heavier p-nuclides, typically above the rare-earth region, the driver pattern becomes more complex. In several cases, the dominant reaction is displaced from the p-nucleus itself and is associated with a neighbouring isotopic chain. Such drivers often appear with a negative correlation sign, indicating that increasing the corresponding rate diverts the flow away from the considered p-nucleus. This behaviour is characteristic of branching regions where $(\gamma,n)$ reactions compete with $(\gamma,p)$ or $(\gamma,\alpha)$ channels and where moderate rate variations can modify the direction of the reaction flow.

The clustering of statistically identified drivers around independently determined branching regions therefore supports the reliability of the method. It shows that the analysis does not merely select reactions with large statistical leverage, but recovers reactions located in physically meaningful regions of the network where the p-process flow is expected to be sensitive to nuclear uncertainties.

\subsection{Comparison across p-process sites}

The same representation can be used to compare the core drivers obtained in the three astrophysical environments considered in this work. Figure~\ref{fig:NZ_triple_site_reactions} shows the distribution of the influential reactions in the $(N,Z)$ plane for the W7 SNIa model, the $25,M_\odot$ model at solar metallicity, and the rotating $25,M_\odot$ model at low metallicity. In contrast to Fig.~\ref{fig:NZ_map_core_drivers_average_vs_top}, where the outer ring indicates the layer-selection method, the ring in Fig.~\ref{fig:NZ_triple_site_reactions} identifies the astrophysical site in which the reaction is selected as a core driver.

\begin{figure*}
\centering
\includegraphics[width=\linewidth, height=0.6\linewidth]{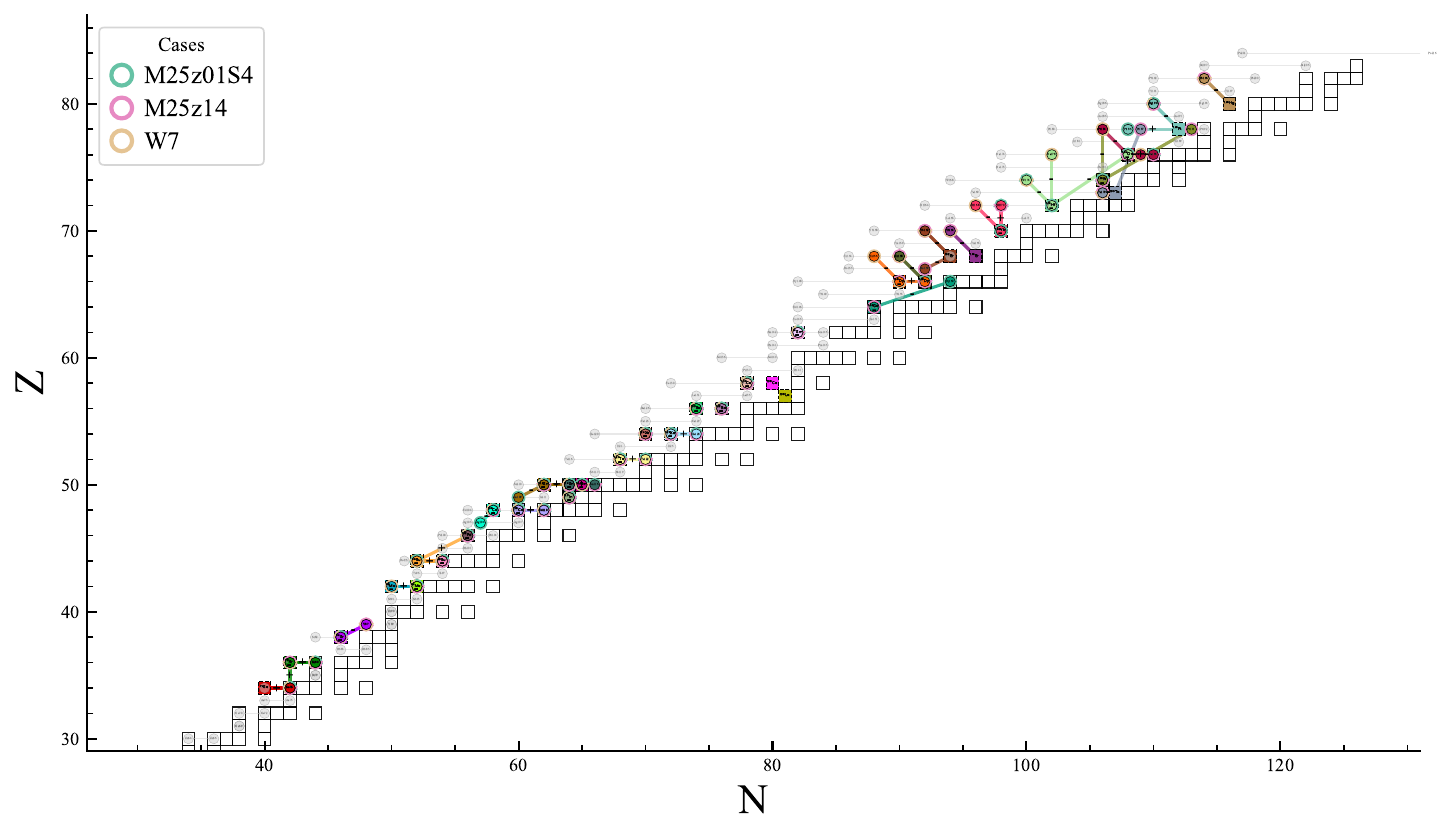}
\caption{
Same graphical representation as in Fig.~\ref{fig:NZ_focus}, but comparing the core drivers obtained for the three astrophysical sites considered in this work: the W7 SNIa model, the $25,M_\odot$ model at solar metallicity (M25z14), and the rotating $25,M_\odot$ model at low metallicity (M25z01S4). The colored outer ring around each driver indicates in which site(s) the reaction is identified as a core driver. An interactive version of the nuclear chart is available online at
\href{https://sebastienmartinet.github.io/NZ_driver_map}{https://sebastienmartinet.github.io/NZ\_driver\_map}.
}

\label{fig:NZ_triple_site_reactions}
\end{figure*}

Many of the dominant reactions are common to more than one site, and in several cases to all three. For example, the uncertainty of $^{74}$Se is controlled by $^{76}$Se$(\gamma,n)$ in all environments considered here. Such reactions are particularly valuable experimental targets, since reducing their uncertainties would improve the reliability of p-process predictions in several astrophysical scenarios simultaneously.

At the same time, Fig.~\ref{fig:NZ_triple_site_reactions} also reveals site-dependent differences. These differences reflect the fact that the detailed sensitivity pattern depends not only on the local nuclear flow, but also on the thermodynamic history and seed distribution of each model. Even when the dominant driver remains unchanged, secondary drivers may differ from one site to another.

Despite these differences, the global picture remains robust: for most p-nuclides, the abundance uncertainty is controlled by a small number of local photodisintegration reactions, most often the photoneutron emission of the p-nucleus itself or that of a nearby isotope along the same isotopic chain. More distant drivers appear in some cases, especially for heavier nuclei, where flow redistribution through branching regions becomes more important.

This result has an important practical implication. Since many of the identified drivers correspond to stable or near-stable nuclei located close to the p-nuclides themselves, a substantial fraction of the key reactions should be experimentally accessible, either directly or through inverse or indirect constraints. Although laboratory photoneutron measurements probe only the ground-state contribution to stellar rates, they can still provide valuable constraints on the PSFs. Similarly, indirect approaches such as the Oslo method \citep{Wiedeking24,Wiedeking25} can constrain PSFs and NLDs and thereby reduce the parameter uncertainties affecting the corresponding key reactions. A list of the most impactful photoneutron emissions is given in Table~\ref{tab:table_3_sites}.
}

\section{Conclusions}
\label{sec:conc}

In this work, we investigated the impact of nuclear reaction rate uncertainties on p-process nucleosynthesis in different supernova environments. To this end, we extended the Backward--Forward Monte Carlo (BFMC) approach applied to radiative neutron captures in \citet{Martinet24} to the p-process regime, allowing us to propagate local parameter uncertainties in the NLDs and PSFs to correlated sets of $(n,\gamma)$, $(p,\gamma)$, and $(\alpha,\gamma)$ rates, and consequently to their inverse photodisintegration channels.

Using these correlated extreme rate sets, we propagated the nuclear uncertainties through full p-process network calculations for three representative astrophysical sites: the W7 Type Ia supernova model, a $25\,M_\odot$ massive star at solar metallicity, and a rotating $25\,M_\odot$ model at low metallicity. We found that the resulting abundance uncertainties are of similar magnitude in all three cases, with a typical spread of about $0.7$ dex, corresponding to roughly a factor of five between the minimum and maximum predicted abundances. Although the overall abundance patterns differ from one site to another because of their different seed distributions and thermodynamic histories, the relative impact of the nuclear uncertainties remains broadly comparable.

By isolating the contribution of the different reaction channels, we showed that the uncertainty budget is overwhelmingly dominated by the $(\gamma,n)$ rates, and therefore by the corresponding calculated $(n,\gamma)$ channels. In contrast, the $(p,\gamma)$ and $(\alpha,\gamma)$ channels have only a minor effect on the final abundance spread within the present uncertainty framework associated with NLD and PSF parameter uncertainties. This result justifies our strategy of identifying, for each nucleus, the maximum and minimum $(n,\gamma)$ rates and associating to them the corresponding correlated $(p,\gamma)$ and $(\alpha,\gamma)$ rates computed with the same nuclear input configuration.

A comparison between statistical parameter uncertainties and systematic model uncertainties further showed that, within the class of modern microscopic NLD and PSF models considered here, the correlated model dependence remains small. The dominant source of uncertainty arises instead from the range of local parameter variations still allowed by the present experimental constraints. In this sense, the main limitation is the lack of sufficiently constraining nuclear data in the relevant proton-rich region.

To identify which reactions most strongly control the abundance uncertainties of the p-nuclides, we combined regularized linear-response modeling, stability analysis, and contribution and interaction decompositions. This statistical framework allowed us to isolate a set of core drivers for each p-nucleus and to visualize them directly in the $(N,Z)$ plane. The resulting distributions show that, for a large fraction of the p-nuclides, the dominant reactions are either the photoneutron emission of the p-nucleus itself or the $(\gamma,n)$ reaction on a nearby isotope along the same isotopic chain. The identified drivers are also found to cluster close to independently determined branching regions, providing a strong physical validation of the method.

An especially important outcome of this study is that many of the key reactions identified as dominant drivers correspond to stable or near-stable nuclei, and in several cases to the photodisintegration of the p-nuclei themselves. This means that a substantial fraction of the reactions controlling the p-process abundance uncertainties should, at least in principle, be experimentally accessible through direct measurements. In particular, improving the constraints on the photodisintegration of stable p-nuclei and their closest isotopic neighbours offers a realistic path towards significantly reducing the uncertainty associated with NLDs and PSFs in p-process calculations.

\begin{acknowledgements}
SM and SG received support from the European Union (ChECTEC-INFRA, project Nr 101008324). This work was supported by the F.R.S.-FNRS under Grant Nr IISN 4.4502.19 and by the F.R.S.-FNRS and the Fonds Wetenschappelijk Onderzoek - Vlaanderen (FWO) under the EOS Project Nr O000422. SG is senior F.R.S.-FNRS research associate. A.C. is post-doctorate F.R.S-FNRS fellow. The present research benefited from computational resources made available on Lucia, the Tier-1 supercomputer of the Walloon Region, infrastructure funded by the Walloon Region under the grant agreement Nr 1910247. SM, SG and AC are member of BLU-ULB, the interfaculty research group focusing on space research at ULB - Université libre de Bruxelles.
\end{acknowledgements}

\bibliographystyle{aa} % style aa.bst
\bibliography{astro} % your references Yourfile.bib

\clearpage
\appendix

\section{Nuclear parameter uncertainties in the correlated $(\alpha,\gamma)$ and $(p,\gamma)$ channels}

In addition to the $(n,\gamma)$ uncertainties shown in Fig.~\ref{fig:NZ_max_min_rate_p_pro_n}, we present here the corresponding correlated uncertainties for the $(\alpha,\gamma)$ and $(p,\gamma)$ reaction rates. These rates are not varied independently. Instead, for each nucleus, we identify the parameter sets that produce the maximum and minimum $(n,\gamma)$ rates within the BFMC framework, and we then extract the associated $(\alpha,\gamma)$ and $(p,\gamma)$ rates computed with the same nuclear input configuration. In this way, the figures shown here represent the correlated variations of the charged-particle capture channels associated with the extrema of the neutron-capture rates.

\begin{figure*}
    \centering
    \includegraphics[width=0.99\textwidth]{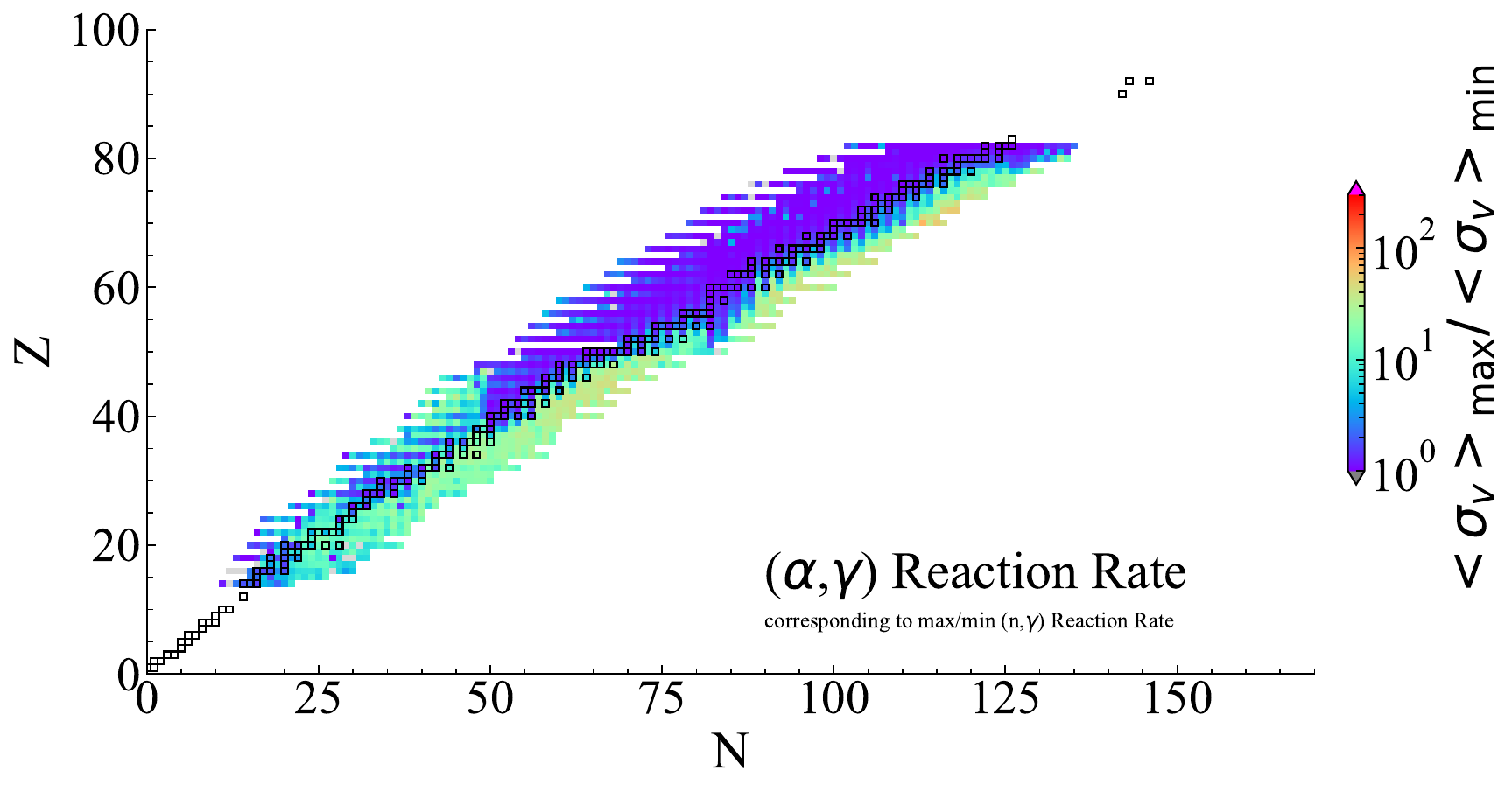}
    \caption{Representation in the ($N$,$Z$) plane of the nuclear parameter uncertainties on the correlated $(\alpha,\gamma)$ reaction rates associated with the extrema of the $(n,\gamma)$ rates. The color code depicts the ratio between the maximum and minimum correlated $(\alpha,\gamma)$ rates.}
    \label{fig:NZ_max_min_rate_p_pro_a_sub}
\end{figure*}

Figure~\ref{fig:NZ_max_min_rate_p_pro_a_sub} shows the resulting uncertainty range for the correlated $(\alpha,\gamma)$ rates. The largest variations are found on the neutron-rich side of the chart, while on the proton-rich side the uncertainties remain generally very small, except for a limited number of light nuclei. This behaviour reflects the fact that the NLDs and PSFs have only a limited impact on $\alpha$-capture rates in the proton-rich region relevant to the p-process, the main ingredient within the Gamow energy range of interest being the $\alpha$-nucleus optical model potential.

\begin{figure*}
    \centering
    \includegraphics[width=0.99\textwidth]{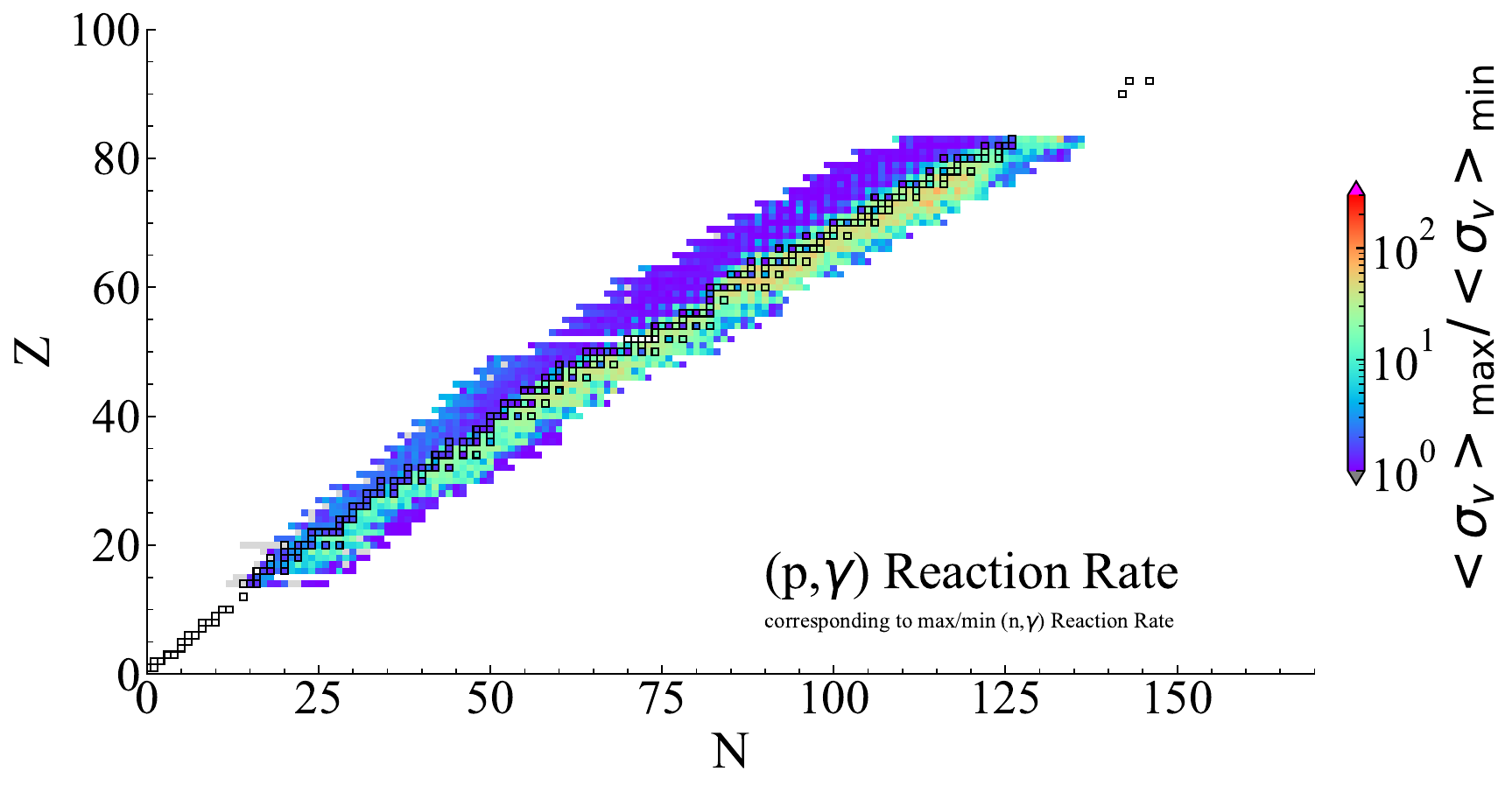}
    \caption{Same as Fig.~\ref{fig:NZ_max_min_rate_p_pro_a_sub}, but for the correlated $(p,\gamma)$ rates associated with the extrema of the $(n,\gamma)$ rates. The color code depicts the ratio between the maximum and minimum correlated $(p,\gamma)$ rates.}
    \label{fig:NZ_max_min_rate_p_pro_p_sub}
\end{figure*}

Figure~\ref{fig:NZ_max_min_rate_p_pro_p_sub} also shows the corresponding result for the correlated $(p,\gamma)$ rates. On the proton-rich side of the chart, the uncertainties remain very limited, whereas most of the residual sensitivity to the NLD and PSF inputs is concentrated on the neutron-rich side, especially near the valley of stability and for heavier nuclei.

These two figures therefore explain, at the level of the reaction rate uncertainties themselves, why the NLD and PSF uncertainties affecting the isolated $(p,\gamma)$ and $(\alpha,\gamma)$ channels have only a minor impact on the final p-process abundances, as discussed in Sec.~\ref{sec:isolimp}. Within the present framework focusing on the parameter uncertainties related to NLDs and PSFs, the dominant contribution clearly originates from the $(n,\gamma)$ channel and its inverse photoneutron emission rates.

\section{Methodology for determining important reactions}
\label{app:ml_details}

This appendix details the statistical workflow used to identify the reaction rates driving the abundance uncertainties of the p-nuclei. The goal of the procedure is not to build a predictive surrogate model of the full nucleosynthesis network, but to identify, for each target p-nucleus, the reactions that most strongly control the abundance variations induced by the Monte Carlo rate sets.

For each target p-nucleus, the abundance obtained in each Monte Carlo realization is associated with the corresponding set of reaction-rate variations used in that simulation. The input features are therefore the adopted rate states, corresponding to the minimum or maximum rates selected for each reaction, while the response variable is the final abundance of the target nucleus.

\subsection{Analysis windows and robustness criterion}

For each target p-nucleus, the statistical analysis was restricted to a local window in the $(N,Z)$ plane centred on the target nucleus. The purpose of this window is to define the subset of neighbouring reactions used as candidate predictors for the abundance variation of that nucleus. This restriction reduces the dimensionality of the regression problem and reflects the fact that the abundance of a p-nucleus is expected to be controlled primarily by reactions located in its local nuclear neighbourhood.

The complete ElasticNet--Random-Forest--SHAP procedure was repeated for seven different window sizes. The smallest windows include only reactions close to the target nucleus, while progressively larger windows include increasingly distant reactions. The largest window approaches the full reaction network considered in the statistical analysis. The final classification is therefore not tied to a single arbitrary choice of window size.

To avoid introducing spurious statistical correlations from reactions located far below the target isotopic chain, the windows were restricted on the low-$Z$ side. This choice keeps the neighbouring isotopic chains that may contribute to the local photodisintegration flow, while excluding distant lower-$Z$ reactions that are not physically expected to influence the abundance of the target p-nucleus.

A reaction was retained as a robust driver only if it was recovered consistently across at least six of the seven window sizes. This robustness criterion reduces the probability of selecting reactions that appear important only because of a particular window definition or because of local statistical fluctuations. The final list of core drivers therefore contains reactions that are both statistically important and persistent under changes in the local analysis domain.

\subsection{Linear screening and dimensionality reduction}

For each target nucleus and each analysis window, the reaction-rate assignments constitute the input features, while the final abundance of the target nucleus forms the response variable.

An ElasticNet regression model was first applied as a dimensionality-reduction step. The purpose of this stage is not to identify the final important reaction rates, but only to remove reactions exhibiting negligible influence on the abundance variations before applying the non-linear analysis. The final driver classification is performed at a later stage using the Random Forest and SHAP analyses described below.

The ElasticNet model was fitted using five-fold cross validation and three values of the mixing parameter,
\begin{equation}
l_1 = {0.1,,0.5,,0.9},
\end{equation}
allowing the optimal regularization strength and mixing ratio to be determined automatically. All input features were standardized prior to fitting.

To assess coefficient stability, the fitting procedure was repeated over the five cross-validation folds. A reaction rate was considered selected within a fold when the absolute value of its coefficient exceeded
\begin{equation}
|\beta| > 10^{-3}.
\end{equation}
The selection frequency was then computed as the fraction of folds in which the coefficient remained non-zero.

In practice, this dimensionality-reduction stage was deliberately conservative. No hard stability threshold was imposed at this stage, and only the weakest retained non-zero coefficients were removed when the number of candidate reactions remained large. More specifically, when more than thirty candidate reactions were retained, the weakest $5\%$ of the non-zero coefficients were discarded. This procedure removes only reactions with negligible statistical leverage, while preserving the physically relevant uncertainty structure for the subsequent non-linear analysis.

\subsection{Random Forest model}

The reduced set of candidate reaction rates was subsequently analysed using a Random Forest regressor composed of 100 decision trees. Unlike the ElasticNet approximation, the Random Forest model can capture non-linear dependencies between reaction rates and final abundances. It is therefore better suited to describing the coupled behaviour of a nuclear reaction network, where the impact of one reaction rate may depend on the values adopted for other rates.

The Random Forest model was not used as a purely predictive tool. Instead, it provided a non-linear representation of the relationship between the retained reaction-rate variations and the abundance response of each target p-nucleus. This representation was then interpreted using SHAP values, as described below.

\subsection{SHAP main effects and interaction effects}

To quantify the contribution of each reaction rate to the abundance variations, Shapley values were computed using the TreeSHAP formalism. Shapley values provide an additive decomposition of the model prediction into contributions associated with each input variable. In the present context, they allow us to quantify how much each reaction-rate variation contributes to the abundance variation of a given p-nucleus.

For each reaction rate $i$, the main contribution was defined as
\begin{equation}
M_i = \left< |\phi_i| \right>,
\end{equation}
where $\phi_i$ is the SHAP value associated with reaction rate $i$, and the brackets indicate an average over all Monte Carlo realizations. The quantity $M_i$ therefore measures the average direct contribution of reaction rate $i$ to the abundance variation.

To quantify coupled effects between reaction rates, SHAP interaction values were also computed. The total interaction strength of reaction rate $i$ was defined as
\begin{equation}
I_i =
\sum_j
\left< |\phi_{ij}| \right>
- 
\left< |\phi_i| \right>,
\end{equation}
where $\phi_{ij}$ is the SHAP interaction contribution between reaction rates $i$ and $j$. The quantity $I_i$ therefore measures the extent to which the influence of reaction rate $i$ depends on the values adopted for other reaction rates in the network.

The two quantities $M_i$ and $I_i$ provide complementary information. The main effect $M_i$ measures the direct importance of a reaction rate, while the interaction strength $I_i$ measures its participation in coupled network behaviour. A reaction with a large main effect but weak interaction strength acts primarily through an individual abundance pathway. Conversely, a reaction with a strong interaction contribution may not dominate alone but can become important through its coupling with other reactions. Reactions that are large in both quantities are therefore interpreted as the most important drivers of abundance uncertainty.

\subsection{Classification of reaction rates}

For each target nucleus and analysis window, the main-effect and interaction-strength distributions were normalized to the interval $[0,1]$. Reaction rates were then assigned to different categories according to their normalized values of $M_i$ and $I_i$:
\begin{itemize}
\item \textit{Core Drivers}: $M_i \geq 0.8$ and $I_i \geq 0.8$;
\item \textit{High Main}: $M_i \geq 0.5$ and $I_i < 0.8$;
\item \textit{High Interaction}: $I_i \geq 0.5$ and $M_i < 0.8$;
\item \textit{Moderate}: $M_i \geq 0.2$ or $I_i \geq 0.2$;
\item \textit{Weak}: all remaining reactions.
\end{itemize}

The \textit{Core Driver} category therefore selects reactions that simultaneously exhibit a strong direct contribution and a strong interaction contribution. These reactions are interpreted as the dominant drivers of abundance uncertainty because they both affect the abundance individually and participate in the coupled response of the reaction network.

The distinction between the different categories is illustrated in Fig.~\ref{fig:shap_quadrant_category_stats}, which shows the distribution of all reaction rates in the normalized main-effect versus interaction-strength plane. The core-driver population occupies the upper-right region of this diagram, while weak contributors are concentrated at low values of both metrics. This separation provides a statistical validation of the adopted classification.

\subsection{Layer selection}

Two complementary abundance datasets were considered when applying the driver-identification procedure. In the first approach, the abundance of each p-nucleus was averaged over all mass zones contributing to the p-process production. This ``averaged-layer'' analysis provides a global view of the integrated ejecta composition and is therefore representative of the final abundance pattern.

In the second approach, the analysis was restricted to the single layer in which the production of the considered p-nucleus is maximal. This ``top-layer'' analysis isolates the thermodynamic conditions most favourable to the production of that isotope. Comparing the averaged-layer and top-layer results allows us to distinguish reactions that control the global ejecta-integrated abundance from those that are important only under the most favourable local production conditions.

Drivers recovered in both approaches are interpreted as particularly robust. Drivers recovered only in the top-layer analysis may reflect sensitivities that are important locally but have a smaller effect on the final integrated abundance. Conversely, drivers recovered in the averaged-layer analysis are more representative of the global nucleosynthesis outcome.

\subsection{Summary of the workflow}

The complete procedure can therefore be summarized as follows:
\begin{enumerate}
\item for each target p-nucleus, define a local analysis window in the $(N,Z)$ plane;
\item use ElasticNet regression as a conservative screening step to discard reactions with negligible influence;
\item analyse the retained reactions with a Random Forest regressor;
\item compute SHAP main effects and SHAP interaction strengths;
\item classify the reactions according to their normalized direct and interaction contributions;
\item repeat the analysis for seven window sizes;
\item retain as robust core drivers only the reactions consistently recovered across the different windows.
\end{enumerate}

This workflow combines conservative dimensionality reduction, non-linear modelling, interaction decomposition, and robustness testing. It is designed to identify reactions that are not only statistically associated with abundance variations, but also persistent across different local definitions of the nuclear region surrounding each p-nucleus.

\section{List of the most impactful photoneutron emissions}
Table \ref{table:p nuclei} summarises the dominant $(\gamma,n)$ driver rates associated with the abundance uncertainties of the selected $p$-nuclei. For each nucleus, the range of predicted abundances across the Monte Carlo ensemble is reported as $X/X_\odot$, together with the corresponding total uncertainty span, defined as the ratio between the maximum and minimum abundance obtained over all realizations.
For each candidate driver rate, its impact is estimated by separating the Monte Carlo realizations into two subsets: those in which the rate is sampled at its maximum value and those in which it is sampled at its minimum value, while all other rates continue to vary. The mean abundance is computed in each subset and the effect reported in the table is the ratio between these two mean values, i.e. $\langle X \rangle_{\mathrm{max}} / \langle X \rangle_{\mathrm{min}}$. This quantity therefore reflects how the average abundance differs between simulations where the rate takes its upper or lower value.
To quantify the importance of each rate relative to the overall uncertainty, we compute its contribution to the total uncertainty range in logarithmic space. This is defined as the absolute logarithmic difference between the two mean abundances, normalized by the logarithmic width of the full abundance range of the nucleus. It provides a measure of how large the effect associated with a given rate is compared to the total spread obtained in the Monte Carlo calculations.
The column ``Detected in'' indicates whether a given rate is identified in one or multiple filtering configurations. Only rates that are consistently recovered across multiple analysis windows are retained in the table.

The same analysis is repeated for our three different SN models (W7, M25z14, and M25z01S4) in Table \ref{tab:table_3_sites}. For each site, the driver rates are identified using the same procedure, and the corresponding abundance statistics are computed independently. This enables a direct comparison of the dominant $(\gamma,n)$ drivers across different explosion conditions, and allows us to assess which reaction rates systematically control the production of $p$-nuclei across models and which are site-dependent.

\clearpage
\onecolumn
\begin{table}[]
\caption{
Summary of the dominant $(\gamma,n)$ driver rates for each $p$-nucleus in the W7 SNIa model.
Columns are defined as follows:
\textit{p-nucleus} gives the isotope considered;
\textit{$X/X_\odot$ range} is the minimum and maximum predicted abundances relative to solar across the ensemble;
\textit{Ratio span} is the ratio between maximum and minimum abundances;
\textit{Driver} indicates the reaction rate identified as a key contributor;
\textit{Rate ratio} is the uncertainty factor applied to that reaction rate;
\textit{Effect (high vs low)} gives the ratio between the mean abundance obtained in simulations where the rate is sampled at its maximum value and that obtained when it is sampled at its minimum value. Values of the form $\times a$ indicate that the abundance is larger by a factor $a$ in the upper-value subset, while values of the form $\div a$ indicate that it is smaller by the factor $a$.
\textit{Contribution to total
uncertainty range (\%)} is the fraction of the total abundance uncertainty span accounted for by this rate, computed in logarithmic space;
\textit{Detected in} indicates whether the rate is identified in one ``Top layer'' or multiple ``Average Layer''  analysis cases.
The table is available as an interactive version online and can be downloaded in CSV format at
\href{https://sebastienmartinet.github.io/NZ_driver_map}{https://sebastienmartinet.github.io/NZ\_driver\_map}.}

\label{table:p nuclei}
\begin{center}
\resizebox{\textwidth}{!}{
\begin{tabular}{|l|c|c|c|c|c|c|c|}
\toprule
 p-nucleus & $X/X_\odot$ range &  Ratio span &                 Driver &  Rate ratio & Effect&  Contribution to total  &   Detected in \\ & & & & increase &  (high / low)& uncertainty range (\%) &\\
\midrule
 $^{74}$Se &         15.1–55.6 &        3.7 &  $^{76}$Se$(\gamma,n)$ &        12.7 &           $\times 1.88$ &                 48.4 &          Both \\
\hline
 $^{78}$Kr &          4.2–24.0 &        5.7 &  $^{78}$Kr$(\gamma,n)$ &        11.8 &                $\div 1.07$ &                  3.8 &          Both \\
           &                   &            &  $^{80}$Kr$(\gamma,n)$ &        14.6 &           $\times 2.19$ &                 44.7 &          Both \\
\hline
 $^{84}$Sr &         23.2–84.0 &        3.6 &  $^{84}$Sr$(\gamma,n)$ &        11.9 &                $\div 1.38$ &                 24.8 &          Both \\
           &                   &            &   $^{87}$Y$(\gamma,n)$ &        15.0 &                $\div 1.25$ &                 17.5 & Average Layer \\
\hline
 $^{92}$Mo &         18.2–37.4 &        2.1 &  $^{92}$Mo$(\gamma,n)$ &        17.7 &                $\div 1.28$ &                 34.7 & Average Layer \\
           &                   &            &  $^{94}$Mo$(\gamma,n)$ &        11.1 &           $\times 1.30$ &                 36.8 & Average Layer \\
           &                   &            & $^{102}$Pd$(\gamma,n)$ &        14.0 &           $\times 1.23$ &                 28.5 &    Top Layer \\
\hline
 $^{94}$Mo &          8.4–21.8 &        2.6 &  $^{94}$Mo$(\gamma,n)$ &        11.1 &                $\div 2.12$ &                 79.0 &          Both \\
\hline
 $^{96}$Ru &         10.8–69.3 &        6.4 &  $^{96}$Ru$(\gamma,n)$ &        10.8 &                $\div 1.67$ &                 27.5 & Average Layer \\
           &                   &            &  $^{98}$Ru$(\gamma,n)$ &        12.5 &           $\times 2.04$ &                 38.3 & Average Layer \\
\hline
 $^{98}$Ru &        41.9–183.0 &        4.4 &  $^{98}$Ru$(\gamma,n)$ &        12.5 &                $\div 2.48$ &                 61.7 &          Both \\
\hline
$^{102}$Pd &      134.0–1170.0 &        8.7 & $^{102}$Pd$(\gamma,n)$ &        14.0 &                $\div 2.30$ &                 38.4 &          Both \\
\hline
$^{106}$Cd &       157.0–896.0 &        5.7 & $^{106}$Cd$(\gamma,n)$ &        13.0 &                $\div 1.98$ &                 39.1 &          Both \\
\hline
$^{108}$Cd &        81.0–588.0 &        7.3 & $^{108}$Cd$(\gamma,n)$ &        10.4 &                $\div 3.10$ &                 57.0 &          Both \\
           &                   &            & $^{110}$Cd$(\gamma,n)$ &        10.9 &           $\times 1.83$ &                 30.5 & Average Layer \\
\hline
$^{113}$In &          7.1–63.1 &        8.9 & $^{113}$In$(\gamma,n)$ &        18.4 &                $\div 2.54$ &                 42.6 &          Both \\
\hline
$^{112}$Sn &        83.2–637.0 &        7.7 & $^{112}$Sn$(\gamma,n)$ &        10.5 &                $\div 1.87$ &                 30.8 &          Both \\
           &                   &            & $^{114}$Sn$(\gamma,n)$ &        13.9 &           $\times 1.27$ &                 11.7 & Average Layer \\
\hline
$^{114}$Sn &       147.0–847.0 &        5.8 & $^{114}$Sn$(\gamma,n)$ &        13.9 &                $\div 2.59$ &                 54.3 &          Both \\
\hline
$^{115}$Sn &           0.4–7.5 &       19.7 & $^{115}$Sn$(\gamma,n)$ &         6.5 &                $\div 5.32$ &                 56.1 &          Both \\
\hline
$^{120}$Te &      291.0–1530.0 &        5.3 & $^{120}$Te$(\gamma,n)$ &        11.2 &                $\div 2.89$ &                 63.9 &          Both \\
\hline
$^{124}$Xe &      239.0–1320.0 &        5.5 & $^{124}$Xe$(\gamma,n)$ &         6.2 &                $\div 2.57$ &                 55.3 &          Both \\
\hline
$^{126}$Xe &      358.0–1320.0 &        3.7 & $^{126}$Xe$(\gamma,n)$ &         8.0 &                $\div 1.52$ &                 32.0 & Average Layer \\
\hline
$^{130}$Ba &      424.0–2720.0 &        6.4 & $^{130}$Ba$(\gamma,n)$ &         8.5 &                $\div 2.61$ &                 51.6 &          Both \\
\hline
$^{132}$Ba &      326.0–1700.0 &        5.2 & $^{132}$Ba$(\gamma,n)$ &         7.4 &                $\div 2.06$ &                 43.8 &          Both \\
\hline
$^{138}$La &          1.1–25.7 &       22.9 & $^{138}$La$(\gamma,n)$ &        12.1 &                $\div 2.01$ &                 22.2 &    Top Layer \\
\hline
$^{136}$Ce &       180.0–875.0 &        4.9 & $^{136}$Ce$(\gamma,n)$ &         5.4 &                $\div 1.90$ &                 40.6 & Average Layer \\
\hline
$^{138}$Ce &       183.0–919.0 &        5.0 & $^{138}$Ce$(\gamma,n)$ &         8.2 &                $\div 1.54$ &                 26.9 &    Top Layer \\
\hline
$^{144}$Sm &       282.0–642.0 &        2.3 & $^{144}$Sm$(\gamma,n)$ &        10.7 &                $\div 1.78$ &                 69.8 &          Both \\
\hline
$^{152}$Gd &         15.7–40.5 &        2.6 & $^{152}$Gd$(\gamma,n)$ &        12.7 &                $\div 1.35$ &                 31.7 &    Top Layer \\
           &                   &            & $^{154}$Gd$(\gamma,n)$ &         9.8 &           $\times 1.08$ &                  7.8 &    Top Layer \\
\hline
$^{156}$Dy &      538.0–1030.0 &        1.9 & $^{156}$Dy$(\gamma,n)$ &         5.2 &                $\div 1.22$ &                 31.0 & Average Layer \\
           &                   &            & $^{156}$Er$(\gamma,n)$ &         3.6 &                $\div 1.25$ &                 34.2 &          Both \\
\hline
$^{158}$Dy &       244.0–596.0 &        2.4 & $^{158}$Er$(\gamma,n)$ &         2.4 &                $\div 1.54$ &                 48.1 &          Both \\
\hline
$^{162}$Er &     1010.0–1470.0 &        1.5 & $^{162}$Yb$(\gamma,n)$ &         1.6 &                $\div 1.21$ &                 50.3 &          Both \\
\hline
$^{164}$Er &         26.9–50.3 &        1.9 & $^{164}$Yb$(\gamma,n)$ &         2.6 &                $\div 1.55$ &                 70.5 &          Both \\
           &                   &            & $^{177}$Os$(\gamma,n)$ &         2.0 &                $\div 1.14$ &                 20.3 &    Top Layer \\
\hline
$^{168}$Yb &     1130.0–1850.0 &        1.6 & $^{168}$Hf$(\gamma,n)$ &         2.1 &                $\div 1.33$ &                 58.2 &          Both \\
\hline
$^{174}$Hf &      592.0–1300.0 &        2.2 &  $^{174}$W$(\gamma,n)$ &         2.9 &                $\div 1.42$ &                 44.4 &          Both \\
           &                   &            & $^{178}$Os$(\gamma,n)$ &         2.0 &                $\div 1.17$ &                 20.1 &          Both \\
\hline
$^{180}$Ta &         23.0–29.5 &        1.3 & $^{179}$Ta$(\gamma,n)$ &        12.1 &                $\div 1.19$ &                 69.5 &          Both \\
\hline
 $^{180}$W &      539.0–1370.0 &        2.5 & $^{180}$Os$(\gamma,n)$ &         2.4 &                $\div 1.30$ &                 28.5 &    Top Layer \\
           &                   &            & $^{184}$Pt$(\gamma,n)$ &         3.9 &                $\div 1.29$ &                 27.4 & Average Layer \\
\hline
$^{184}$Os &     1000.0–2110.0 &        2.1 & $^{178}$Os$(\gamma,n)$ &         2.0 &           $\times 1.21$ &                 25.4 &    Top Layer \\
           &                   &            & $^{184}$Pt$(\gamma,n)$ &         3.9 &                $\div 1.50$ &                 54.4 &          Both \\
           &                   &            & $^{185}$Os$(\gamma,n)$ &         8.8 &                $\div 1.22$ &                 26.4 & Average Layer \\
           &                   &            & $^{188}$Hg$(\gamma,n)$ &         1.8 &                $\div 1.15$ &                 18.5 &    Top Layer \\
\hline
$^{190}$Pt &      969.0–1810.0 &        1.9 & $^{190}$Hg$(\gamma,n)$ &         2.4 &                $\div 1.56$ &                 71.7 &          Both \\
\hline
$^{196}$Hg &       466.0–675.0 &        1.4 & $^{196}$Pb$(\gamma,n)$ &         1.3 &                $\div 1.18$ &                 44.4 & Average Layer \\
           &                   &            & $^{198}$Pb$(\gamma,n)$ &         1.6 &           $\times 1.01$ &                  2.7 &    Top Layer \\
           &                   &            & $^{199}$Pb$(\gamma,n)$ &         2.5 &                $\div 1.01$ &                  1.9 &    Top Layer \\
\bottomrule
\end{tabular}
}
\end{center}
\end{table}

\twocolumn
\clearpage

\onecolumn
\begin{longtable}{|l|c|c|c|l|c|c|c|}
\caption{
Same as Table \ref{table:p nuclei}, but computed for the three SN models (W7, M25z14, and M25z01S4) for the ``averaged layers'' case. For each $p$-nucleus and each site, the table lists the dominant $(\gamma,n)$ driver rates and their associated impact on the abundance.
}
\label{tab:table_3_sites}
\\
\toprule
 p-nucleus &                   Site & $X/X_\odot$ range & Ratio span &                 Driver & Rate ratio &        Effect & Contribution (\%) \\
\midrule
\endfirsthead

\toprule
 p-nucleus &                   Site & $X/X_\odot$ range & Ratio span &                 Driver & Rate ratio &        Effect & Contribution (\%) \\
\midrule
\endhead
\midrule
\multicolumn{8}{r}{{Continued on next page}} \\
\midrule
\endfoot

\bottomrule
\endlastfoot
 $^{74}$Se & \hspace{0.2cm}M25z01S4 &       84.3--167.0 &        2.0 &  $^{76}$Se$(\gamma,n)$ &       12.7 &  $\times 1.2$ &              28.3 \\
           &   \hspace{0.2cm}M25z14 &      175.0--530.0 &        3.0 &  $^{76}$Se$(\gamma,n)$ &       12.7 &  $\times 1.7$ &              50.4 \\
           &       \hspace{0.2cm}W7 &        15.1--55.6 &        3.7 &  $^{76}$Se$(\gamma,n)$ &       12.7 &  $\times 1.9$ &              48.4 \\
\addlinespace[3pt]
\hline \addlinespace[3pt]
 $^{78}$Kr & \hspace{0.2cm}M25z01S4 &       55.2--114.0 &        2.1 &  $^{76}$Se$(\gamma,n)$ &       12.7 &  $\times 1.2$ &              23.4 \\
           &                        &                   &            &  $^{80}$Kr$(\gamma,n)$ &       14.6 &  $\times 1.4$ &              44.7 \\
           &   \hspace{0.2cm}M25z14 &       24.3--151.0 &        6.2 &  $^{78}$Kr$(\gamma,n)$ &       11.8 &  $\times 1.0$ &               0.1 \\
           &                        &                   &            &  $^{80}$Kr$(\gamma,n)$ &       14.6 &  $\times 2.6$ &              51.3 \\
           &       \hspace{0.2cm}W7 &         4.2--24.0 &        5.7 &  $^{78}$Kr$(\gamma,n)$ &       11.8 &    $\div 1.1$ &               3.8 \\
           &                        &                   &            &  $^{80}$Kr$(\gamma,n)$ &       14.6 &  $\times 2.2$ &              44.7 \\
\addlinespace[3pt]
\hline \addlinespace[3pt]
 $^{84}$Sr & \hspace{0.2cm}M25z01S4 &      208.0--335.0 &        1.6 &  $^{84}$Sr$(\gamma,n)$ &       11.9 &    $\div 1.2$ &              41.6 \\
           &   \hspace{0.2cm}M25z14 &       81.9--205.0 &        2.5 &  $^{84}$Sr$(\gamma,n)$ &       11.9 &    $\div 1.3$ &              25.7 \\
           &                        &                   &            &   $^{87}$Y$(\gamma,n)$ &       15.0 &    $\div 1.2$ &              17.8 \\
           &       \hspace{0.2cm}W7 &        23.2--84.0 &        3.6 &  $^{84}$Sr$(\gamma,n)$ &       11.9 &    $\div 1.4$ &              24.8 \\
           &                        &                   &            &   $^{87}$Y$(\gamma,n)$ &       15.0 &    $\div 1.3$ &              17.5 \\
\addlinespace[3pt]
\hline \addlinespace[3pt]
 $^{92}$Mo & \hspace{0.2cm}M25z01S4 &        20.3--36.8 &        1.8 &  $^{94}$Mo$(\gamma,n)$ &       11.1 &  $\times 1.4$ &              51.3 \\
           &   \hspace{0.2cm}M25z14 &          2.1--4.6 &        2.1 &                     -- &            &               &                   \\
           &       \hspace{0.2cm}W7 &        18.2--37.4 &        2.1 &  $^{92}$Mo$(\gamma,n)$ &       17.7 &    $\div 1.3$ &              34.7 \\
           &                        &                   &            &  $^{94}$Mo$(\gamma,n)$ &       11.1 &  $\times 1.3$ &              36.8 \\
\addlinespace[3pt]
\hline \addlinespace[3pt]
 $^{94}$Mo & \hspace{0.2cm}M25z01S4 &        20.6--38.6 &        1.9 &  $^{94}$Mo$(\gamma,n)$ &       11.1 &    $\div 1.7$ &              81.5 \\
           &   \hspace{0.2cm}M25z14 &          3.0--4.8 &        1.6 &  $^{94}$Mo$(\gamma,n)$ &       11.1 &    $\div 1.5$ &              81.9 \\
           &       \hspace{0.2cm}W7 &         8.4--21.8 &        2.6 &  $^{94}$Mo$(\gamma,n)$ &       11.1 &    $\div 2.1$ &              79.0 \\
\addlinespace[3pt]
\hline \addlinespace[3pt]
 $^{96}$Ru & \hspace{0.2cm}M25z01S4 &         9.1--41.0 &        4.5 &  $^{96}$Ru$(\gamma,n)$ &       10.8 &    $\div 1.5$ &              28.2 \\
           &                        &                   &            &  $^{98}$Ru$(\gamma,n)$ &       12.5 &  $\times 2.0$ &              45.6 \\
           &                        &                   &            & $^{102}$Pd$(\gamma,n)$ &       14.0 &  $\times 1.2$ &              13.6 \\
           &   \hspace{0.2cm}M25z14 &          1.3--7.0 &        5.6 &                     -- &            &               &                   \\
           &       \hspace{0.2cm}W7 &        10.8--69.3 &        6.4 &  $^{96}$Ru$(\gamma,n)$ &       10.8 &    $\div 1.7$ &              27.5 \\
           &                        &                   &            &  $^{98}$Ru$(\gamma,n)$ &       12.5 &  $\times 2.0$ &              38.3 \\
\addlinespace[3pt]
\hline \addlinespace[3pt]
 $^{98}$Ru & \hspace{0.2cm}M25z01S4 &       62.3--162.0 &        2.6 &  $^{98}$Ru$(\gamma,n)$ &       12.5 &    $\div 1.8$ &              63.5 \\
           &   \hspace{0.2cm}M25z14 &         6.7--19.5 &        2.9 &  $^{98}$Ru$(\gamma,n)$ &       12.5 &    $\div 2.0$ &              63.9 \\
           &       \hspace{0.2cm}W7 &       41.9--183.0 &        4.4 &  $^{98}$Ru$(\gamma,n)$ &       12.5 &    $\div 2.5$ &              61.7 \\
\addlinespace[3pt]
\hline \addlinespace[3pt]
$^{102}$Pd & \hspace{0.2cm}M25z01S4 &       88.5--572.0 &        6.5 & $^{102}$Pd$(\gamma,n)$ &       14.0 &    $\div 2.1$ &              40.4 \\
           &   \hspace{0.2cm}M25z14 &       12.5--114.0 &        9.1 & $^{102}$Pd$(\gamma,n)$ &       14.0 &    $\div 2.8$ &              45.8 \\
           &       \hspace{0.2cm}W7 &     134.0--1170.0 &        8.7 & $^{102}$Pd$(\gamma,n)$ &       14.0 &    $\div 2.3$ &              38.4 \\
\addlinespace[3pt]
\hline \addlinespace[3pt]
$^{106}$Cd & \hspace{0.2cm}M25z01S4 &       74.1--326.0 &        4.4 & $^{104}$Ag$(\gamma,n)$ &       12.9 &    $\div 1.7$ &              34.1 \\
           &                        &                   &            & $^{106}$Cd$(\gamma,n)$ &       13.0 &    $\div 1.8$ &              40.9 \\
           &   \hspace{0.2cm}M25z14 &        17.0--77.3 &        4.5 & $^{106}$Cd$(\gamma,n)$ &       13.0 &    $\div 2.2$ &              52.5 \\
           &       \hspace{0.2cm}W7 &      157.0--896.0 &        5.7 & $^{106}$Cd$(\gamma,n)$ &       13.0 &    $\div 2.0$ &              39.1 \\
\addlinespace[3pt]
\hline \addlinespace[3pt]
$^{108}$Cd & \hspace{0.2cm}M25z01S4 &       83.5--353.0 &        4.2 & $^{108}$Cd$(\gamma,n)$ &       10.4 &    $\div 2.2$ &              54.2 \\
           &                        &                   &            & $^{110}$Cd$(\gamma,n)$ &       10.9 &  $\times 1.8$ &              42.3 \\
           &   \hspace{0.2cm}M25z14 &        10.4--50.6 &        4.9 & $^{108}$Cd$(\gamma,n)$ &       10.4 &    $\div 2.3$ &              53.7 \\
           &                        &                   &            & $^{110}$Cd$(\gamma,n)$ &       10.9 &  $\times 1.9$ &              40.1 \\
           &       \hspace{0.2cm}W7 &       81.0--588.0 &        7.3 & $^{108}$Cd$(\gamma,n)$ &       10.4 &    $\div 3.1$ &              57.0 \\
           &                        &                   &            & $^{110}$Cd$(\gamma,n)$ &       10.9 &  $\times 1.8$ &              30.5 \\
\addlinespace[3pt]
\hline \addlinespace[3pt]
$^{113}$In & \hspace{0.2cm}M25z01S4 &      134.0--186.0 &        1.4 & $^{113}$In$(\gamma,n)$ &       18.4 &    $\div 1.3$ &              68.4 \\
           &   \hspace{0.2cm}M25z14 &          2.7--7.6 &        2.8 & $^{113}$In$(\gamma,n)$ &       18.4 &    $\div 1.7$ &              48.6 \\
           &       \hspace{0.2cm}W7 &         7.1--63.1 &        8.9 & $^{113}$In$(\gamma,n)$ &       18.4 &    $\div 2.5$ &              42.6 \\
\addlinespace[3pt]
\hline \addlinespace[3pt]
$^{112}$Sn & \hspace{0.2cm}M25z01S4 &       24.1--188.0 &        7.8 & $^{109}$In$(\gamma,n)$ &       14.4 &    $\div 1.3$ &              12.1 \\
           &                        &                   &            & $^{112}$Sn$(\gamma,n)$ &       10.5 &    $\div 1.7$ &              25.2 \\
           &   \hspace{0.2cm}M25z14 &         6.4--46.3 &        7.3 & $^{112}$Sn$(\gamma,n)$ &       10.5 &    $\div 2.3$ &              41.9 \\
           &                        &                   &            & $^{114}$Sn$(\gamma,n)$ &       13.9 &  $\times 1.0$ &               1.5 \\
           &       \hspace{0.2cm}W7 &       83.2--637.0 &        7.7 & $^{112}$Sn$(\gamma,n)$ &       10.5 &    $\div 1.9$ &              30.8 \\
           &                        &                   &            & $^{114}$Sn$(\gamma,n)$ &       13.9 &  $\times 1.3$ &              11.7 \\
\addlinespace[3pt]
\hline \addlinespace[3pt]
$^{114}$Sn & \hspace{0.2cm}M25z01S4 &       71.9--341.0 &        4.7 & $^{112}$Sn$(\gamma,n)$ &       10.5 &    $\div 1.4$ &              21.1 \\
           &                        &                   &            & $^{114}$Sn$(\gamma,n)$ &       13.9 &    $\div 1.9$ &              39.5 \\
           &   \hspace{0.2cm}M25z14 &        10.6--58.2 &        5.5 & $^{114}$Sn$(\gamma,n)$ &       13.9 &    $\div 2.2$ &              45.7 \\
           &                        &                   &            & $^{116}$Sn$(\gamma,n)$ &       11.2 &  $\times 1.2$ &               8.5 \\
           &       \hspace{0.2cm}W7 &      147.0--847.0 &        5.8 & $^{114}$Sn$(\gamma,n)$ &       13.9 &    $\div 2.6$ &              54.3 \\
\addlinespace[3pt]
\hline \addlinespace[3pt]
$^{115}$Sn & \hspace{0.2cm}M25z01S4 &         8.6--13.5 &        1.6 & $^{115}$Sn$(\gamma,n)$ &        6.5 &    $\div 1.3$ &              52.4 \\
           &   \hspace{0.2cm}M25z14 &          1.1--1.9 &        1.7 &                     -- &            &               &                   \\
           &       \hspace{0.2cm}W7 &          0.4--7.5 &       19.7 & $^{115}$Sn$(\gamma,n)$ &        6.5 &    $\div 5.3$ &              56.1 \\
\addlinespace[3pt]
\hline \addlinespace[3pt]
$^{120}$Te & \hspace{0.2cm}M25z01S4 &       76.5--332.0 &        4.3 & $^{120}$Te$(\gamma,n)$ &       11.2 &    $\div 2.4$ &              58.5 \\
           &   \hspace{0.2cm}M25z14 &        17.3--88.3 &        5.1 & $^{120}$Te$(\gamma,n)$ &       11.2 &    $\div 2.6$ &              58.1 \\
           &                        &                   &            & $^{122}$Te$(\gamma,n)$ &        7.6 &  $\times 1.8$ &              37.5 \\
           &       \hspace{0.2cm}W7 &     291.0--1530.0 &        5.3 & $^{120}$Te$(\gamma,n)$ &       11.2 &    $\div 2.9$ &              63.9 \\
\addlinespace[3pt]
\hline \addlinespace[3pt]
$^{124}$Xe & \hspace{0.2cm}M25z01S4 &       27.5--145.0 &        5.3 & $^{124}$Xe$(\gamma,n)$ &        6.2 &    $\div 2.2$ &              46.6 \\
           &   \hspace{0.2cm}M25z14 &        14.5--77.3 &        5.3 & $^{124}$Xe$(\gamma,n)$ &        6.2 &    $\div 2.4$ &              52.7 \\
           &       \hspace{0.2cm}W7 &     239.0--1320.0 &        5.5 & $^{124}$Xe$(\gamma,n)$ &        6.2 &    $\div 2.6$ &              55.3 \\
\addlinespace[3pt]
\hline \addlinespace[3pt]
$^{126}$Xe & \hspace{0.2cm}M25z01S4 &       36.2--167.0 &        4.6 & $^{126}$Xe$(\gamma,n)$ &        8.0 &    $\div 1.9$ &              41.1 \\
           &   \hspace{0.2cm}M25z14 &        23.2--85.2 &        3.7 & $^{128}$Xe$(\gamma,n)$ &        9.2 &  $\times 1.5$ &              30.2 \\
           &       \hspace{0.2cm}W7 &     358.0--1320.0 &        3.7 & $^{126}$Xe$(\gamma,n)$ &        8.0 &    $\div 1.5$ &              32.0 \\
\addlinespace[3pt]
\hline \addlinespace[3pt]
$^{130}$Ba & \hspace{0.2cm}M25z01S4 &       29.8--207.0 &        6.9 & $^{130}$Ba$(\gamma,n)$ &        8.5 &    $\div 2.0$ &              36.5 \\
           &   \hspace{0.2cm}M25z14 &       40.1--258.0 &        6.4 & $^{130}$Ba$(\gamma,n)$ &        8.5 &    $\div 2.0$ &              37.9 \\
           &       \hspace{0.2cm}W7 &     424.0--2720.0 &        6.4 & $^{130}$Ba$(\gamma,n)$ &        8.5 &    $\div 2.6$ &              51.6 \\
\addlinespace[3pt]
\hline \addlinespace[3pt]
$^{132}$Ba & \hspace{0.2cm}M25z01S4 &       24.8--166.0 &        6.7 & $^{132}$Ba$(\gamma,n)$ &        7.4 &    $\div 2.0$ &              35.2 \\
           &   \hspace{0.2cm}M25z14 &       44.6--263.0 &        5.9 & $^{132}$Ba$(\gamma,n)$ &        7.4 &    $\div 1.5$ &              23.6 \\
           &       \hspace{0.2cm}W7 &     326.0--1700.0 &        5.2 & $^{132}$Ba$(\gamma,n)$ &        7.4 &    $\div 2.1$ &              43.8 \\
\addlinespace[3pt]
\hline \addlinespace[3pt]
$^{136}$Ce & \hspace{0.2cm}M25z01S4 &         3.4--16.5 &        4.9 & $^{136}$Ce$(\gamma,n)$ &        5.4 &    $\div 1.8$ &              35.9 \\
           &   \hspace{0.2cm}M25z14 &       27.8--135.0 &        4.9 &                     -- &            &               &                   \\
           &       \hspace{0.2cm}W7 &      180.0--875.0 &        4.9 & $^{136}$Ce$(\gamma,n)$ &        5.4 &    $\div 1.9$ &              40.6 \\
\addlinespace[3pt]
\hline \addlinespace[3pt]
$^{144}$Sm & \hspace{0.2cm}M25z01S4 &          3.5--7.3 &        2.1 & $^{144}$Sm$(\gamma,n)$ &       10.7 &    $\div 1.5$ &              54.7 \\
           &   \hspace{0.2cm}M25z14 &       52.8--106.0 &        2.0 & $^{144}$Sm$(\gamma,n)$ &       10.7 &    $\div 1.4$ &              53.1 \\
           &       \hspace{0.2cm}W7 &      282.0--642.0 &        2.3 & $^{144}$Sm$(\gamma,n)$ &       10.7 &    $\div 1.8$ &              69.8 \\
\addlinespace[3pt]
\hline \addlinespace[3pt]
$^{152}$Gd & \hspace{0.2cm}M25z01S4 &        1.0--507.0 &      510.6 & $^{160}$Dy$(\gamma,n)$ &       13.5 & $\times 28.3$ &              53.6 \\
           &   \hspace{0.2cm}M25z14 &        38.6--58.2 &        1.5 & $^{152}$Gd$(\gamma,n)$ &       12.7 &    $\div 1.3$ &              57.4 \\
           &       \hspace{0.2cm}W7 &        15.7--40.5 &        2.6 &                     -- &            &               &                   \\
\addlinespace[3pt]
\hline \addlinespace[3pt]
$^{156}$Dy & \hspace{0.2cm}M25z01S4 &          2.5--8.9 &        3.5 &                     -- &            &               &                   \\
           &   \hspace{0.2cm}M25z14 &       48.5--115.0 &        2.4 & $^{158}$Dy$(\gamma,n)$ &       12.5 &  $\times 1.5$ &              46.6 \\
           &       \hspace{0.2cm}W7 &     538.0--1030.0 &        1.9 & $^{156}$Dy$(\gamma,n)$ &        5.2 &    $\div 1.2$ &              31.0 \\
           &                        &                   &            & $^{156}$Er$(\gamma,n)$ &        3.6 &    $\div 1.2$ &              34.2 \\
\addlinespace[3pt]
\hline \addlinespace[3pt]
$^{158}$Dy & \hspace{0.2cm}M25z01S4 &         5.9--13.6 &        2.3 &                     -- &            &               &                   \\
           &   \hspace{0.2cm}M25z14 &      108.0--213.0 &        2.0 & $^{158}$Er$(\gamma,n)$ &        2.4 &    $\div 1.4$ &              46.7 \\
           &       \hspace{0.2cm}W7 &      244.0--596.0 &        2.4 & $^{158}$Er$(\gamma,n)$ &        2.4 &    $\div 1.5$ &              48.1 \\
\addlinespace[3pt]
\hline \addlinespace[3pt]
$^{162}$Er & \hspace{0.2cm}M25z01S4 &        11.6--20.7 &        1.8 &                     -- &            &               &                   \\
           &   \hspace{0.2cm}M25z14 &      202.0--308.0 &        1.5 & $^{159}$Ho$(\gamma,n)$ &        4.8 &    $\div 1.1$ &              23.7 \\
           &                        &                   &            & $^{162}$Yb$(\gamma,n)$ &        1.6 &    $\div 1.2$ &              37.1 \\
           &       \hspace{0.2cm}W7 &    1010.0--1470.0 &        1.5 & $^{162}$Yb$(\gamma,n)$ &        1.6 &    $\div 1.2$ &              50.3 \\
\addlinespace[3pt]
\hline \addlinespace[3pt]
$^{164}$Er & \hspace{0.2cm}M25z01S4 &          0.5--1.3 &        2.4 &                     -- &            &               &                   \\
           &   \hspace{0.2cm}M25z14 &         9.8--18.9 &        1.9 & $^{164}$Yb$(\gamma,n)$ &        2.6 &    $\div 1.3$ &              38.3 \\
           &       \hspace{0.2cm}W7 &        26.9--50.3 &        1.9 & $^{164}$Yb$(\gamma,n)$ &        2.6 &    $\div 1.6$ &              70.5 \\
\addlinespace[3pt]
\hline \addlinespace[3pt]
$^{168}$Yb & \hspace{0.2cm}M25z01S4 &        23.4--39.4 &        1.7 & $^{168}$Yb$(\gamma,n)$ &        8.1 &    $\div 1.1$ &              26.5 \\
           &   \hspace{0.2cm}M25z14 &      324.0--620.0 &        1.9 & $^{170}$Hf$(\gamma,n)$ &        3.1 &  $\times 1.2$ &              27.0 \\
           &       \hspace{0.2cm}W7 &    1130.0--1850.0 &        1.6 & $^{168}$Hf$(\gamma,n)$ &        2.1 &    $\div 1.3$ &              58.2 \\
\addlinespace[3pt]
\hline \addlinespace[3pt]
$^{174}$Hf & \hspace{0.2cm}M25z01S4 &        10.4--27.9 &        2.7 & $^{174}$Hf$(\gamma,n)$ &        6.9 &    $\div 1.2$ &              15.8 \\
           &   \hspace{0.2cm}M25z14 &      310.0--711.0 &        2.3 & $^{184}$Os$(\gamma,n)$ &        5.4 &    $\div 1.2$ &              18.3 \\
           &       \hspace{0.2cm}W7 &     592.0--1300.0 &        2.2 &  $^{174}$W$(\gamma,n)$ &        2.9 &    $\div 1.4$ &              44.4 \\
           &                        &                   &            & $^{178}$Os$(\gamma,n)$ &        2.0 &    $\div 1.2$ &              20.1 \\
\addlinespace[3pt]
\hline \addlinespace[3pt]
$^{180}$Ta & \hspace{0.2cm}M25z01S4 &          3.7--4.3 &        1.2 &                     -- &            &               &                   \\
           &   \hspace{0.2cm}M25z14 &      118.0--138.0 &        1.2 & $^{179}$Ta$(\gamma,n)$ &       12.1 &    $\div 1.1$ &              53.8 \\
           &                        &                   &            & $^{187}$Pt$(\gamma,n)$ &        2.5 &  $\times 1.0$ &              25.8 \\
           &       \hspace{0.2cm}W7 &        23.0--29.5 &        1.3 & $^{179}$Ta$(\gamma,n)$ &       12.1 &    $\div 1.2$ &              69.5 \\
\addlinespace[3pt]
\hline \addlinespace[3pt]
 $^{180}$W & \hspace{0.2cm}M25z01S4 &        16.3--38.5 &        2.4 &  $^{180}$W$(\gamma,n)$ &       10.2 &    $\div 1.3$ &              32.2 \\
           &   \hspace{0.2cm}M25z14 &     558.0--1090.0 &        2.0 &  $^{180}$W$(\gamma,n)$ &       10.2 &    $\div 1.2$ &              28.5 \\
           &                        &                   &            & $^{191}$Pt$(\gamma,n)$ &        5.2 &    $\div 1.2$ &              23.6 \\
           &       \hspace{0.2cm}W7 &     539.0--1370.0 &        2.5 & $^{184}$Pt$(\gamma,n)$ &        3.9 &    $\div 1.3$ &              27.4 \\
\addlinespace[3pt]
\hline \addlinespace[3pt]
$^{184}$Os & \hspace{0.2cm}M25z01S4 &         7.9--20.1 &        2.5 & $^{184}$Os$(\gamma,n)$ &        5.4 &    $\div 1.3$ &              30.4 \\
           &   \hspace{0.2cm}M25z14 &      381.0--670.0 &        1.8 & $^{186}$Os$(\gamma,n)$ &       10.1 &  $\times 1.3$ &              42.0 \\
           &       \hspace{0.2cm}W7 &    1000.0--2110.0 &        2.1 & $^{184}$Pt$(\gamma,n)$ &        3.9 &    $\div 1.5$ &              54.4 \\
           &                        &                   &            & $^{185}$Os$(\gamma,n)$ &        8.8 &    $\div 1.2$ &              26.4 \\
\addlinespace[3pt]
\hline \addlinespace[3pt]
$^{190}$Pt & \hspace{0.2cm}M25z01S4 &        10.8--52.3 &        4.8 & $^{186}$Pt$(\gamma,n)$ &        3.0 &  $\times 1.2$ &               9.2 \\
           &   \hspace{0.2cm}M25z14 &     635.0--1130.0 &        1.8 & $^{190}$Hg$(\gamma,n)$ &        2.4 &    $\div 1.3$ &              49.7 \\
           &       \hspace{0.2cm}W7 &     969.0--1810.0 &        1.9 & $^{190}$Hg$(\gamma,n)$ &        2.4 &    $\div 1.6$ &              71.7 \\
\addlinespace[3pt]
\hline \addlinespace[3pt]
$^{196}$Hg & \hspace{0.2cm}M25z01S4 &        14.6--31.3 &        2.1 &                     -- &            &               &                   \\
           &   \hspace{0.2cm}M25z14 &      202.0--360.0 &        1.8 & $^{196}$Pb$(\gamma,n)$ &        1.3 &    $\div 1.3$ &              44.8 \\
           &       \hspace{0.2cm}W7 &      466.0--675.0 &        1.4 & $^{196}$Pb$(\gamma,n)$ &        1.3 &    $\div 1.2$ &              44.4 \\
\end{longtable}

\twocolumn
\clearpage

\end{document}